\documentclass[11pt]{article}

\usepackage[margin=1in]{geometry}
\usepackage{setspace}
\usepackage{xeCJK}

\usepackage{amsmath,amssymb,amsfonts,bm,eurosym}

\usepackage{array}
\usepackage{booktabs}
\usepackage{multirow}
\usepackage{tabularx}
\usepackage{longtable}
\usepackage[flushleft]{threeparttable}
\usepackage{colortbl,hhline}
\usepackage{arydshln}

\usepackage{graphicx}
\usepackage{caption}
\usepackage{subcaption}

\usepackage{placeins}
\usepackage{pdflscape}

\usepackage{xcolor}
\usepackage{soul}
\usepackage{ulem}
\usepackage{titlesec}
\usepackage{appendix}
\usepackage{comment}
\usepackage{verbatim}
\usepackage{enumitem}
\usepackage{paracol}
\usepackage{authblk}
\usepackage{footnote}
\usepackage{footmisc}

\usepackage{natbib}

\usepackage{hyperref}
\hypersetup{
    colorlinks=true,
    urlcolor=blue,
    citecolor=[rgb]{0,0.5,0.5}
}
\usepackage{cleveref}

\usepackage{fvextra}

\DefineVerbatimEnvironment{PromptBlock}{Verbatim}{
  breaklines=true,
  breakanywhere=true,
  breaksymbolleft={},
  fontsize=\footnotesize,
  formatcom=\ttfamily,
  baselinestretch=1,
  xleftmargin=0pt
}

\usepackage{tikz}
\usetikzlibrary{arrows.meta,calc,positioning,decorations.pathreplacing}

\definecolor{boxgray}{RGB}{242,240,235}
\definecolor{boxgrayline}{RGB}{190,186,178}

\definecolor{boxpurple}{RGB}{234,231,248}
\definecolor{boxpurpleline}{RGB}{170,165,225}

\definecolor{boxgreen}{RGB}{225,242,237}
\definecolor{boxgreenline}{RGB}{145,200,185}

\definecolor{boxpink}{RGB}{250,236,233}
\definecolor{boxpinkline}{RGB}{224,170,160}

\definecolor{taskblue}{RGB}{238,238,252}
\definecolor{taskblueline}{RGB}{170,165,225}

\definecolor{expyellow}{RGB}{251,241,224}
\definecolor{expyellowline}{RGB}{224,186,130}

\tikzset{
  mybox/.style={
    rectangle,
    rounded corners=4pt,
    minimum width=3.9cm,
    minimum height=1.55cm,
    text width=3.70cm,
    align=center,
    draw=gray!55,
    line width=0.8pt,
    inner sep=4pt,
    font=\small\linespread{0.94}\selectfont
  },
  slimbox/.style={
    rectangle,
    rounded corners=4pt,
    minimum width=3.1cm,
    minimum height=0.80cm,
    text width=3.55cm,
    align=center,
    draw=gray!55,
    line width=0.8pt,
    inner sep=3pt,
    font=\small\linespread{0.94}\selectfont
  },
  arrow/.style={-{Stealth[length=6pt,width=7pt]}, thick, gray!70},
  stageLabel/.style={font=\bfseries\small, anchor=west},
  modelLabel/.style={font=\footnotesize, text=black!75, anchor=west},
  noteFont/.style={font=\footnotesize, text=black!80}
}

\newcolumntype{L}[1]{>{\raggedright\let\newline\\arraybackslash\hspace{0pt}}m{#1}}
\newcolumntype{C}[1]{>{\centering\let\newline\\arraybackslash\hspace{0pt}}m{#1}}
\newcolumntype{R}[1]{>{\raggedleft\let\newline\\arraybackslash\hspace{0pt}}m{#1}}

\usepackage{datetime2}

\usepackage{xkeyval}

\title{\resizebox{\textwidth}{!}{\textbf{Generative AI and the Reorganization of Labor Demand}}}

\author{Fangyan Wang\thanks{Zaiyan Wei gratefully acknowledges research support from the David and Margaret Crow Rising Star Professorship, and Yang Wang gratefully acknowledges financial support from the Mitch Daniels School of Business at Purdue University. We thank Lina Rivas for excellent research assistance. We are also grateful to seminar participants at the University of Delaware for helpful comments and suggestions. All analyses and reported results have been reviewed to ensure that no confidential information is disclosed. Any errors are our own.}}
\author{Zaiyan Wei}
\author{Yang Wang}
\affil{Mitch Daniels School of Business, Purdue University; West Lafayette, IN 47907 \\
    wang6123@purdue.edu; zaiyan@purdue.edu; yangwang@purdue.edu
}

\date{\ifcase\month\or January\or February\or March\or April\or May\or June\or July\or August\or September\or October\or November\or December\fi\ \the\year}

\begin{document}

\begin{titlepage}
    \maketitle

    \vspace{-15pt}

    \begin{abstract}

    \noindent
    Generative artificial intelligence (AI) is expected to transform work, but less is known about how firms reorganize labor demand as the technology diffuses. Existing research has largely focused on which occupations are exposed to AI or whether exposed jobs decline. We extend this debate by examining whether firms adjust by changing where they hire, what jobs contain, or both. Using a nationwide dataset of job postings in the United States, covering all sectors of the economy, we construct a dynamic, posting-level measure of generative AI exposure with a two-stage large language model pipeline. The pipeline identifies the tasks described in each posting and classifies the extent to which generative AI can perform or assist them. We then decompose changes in aggregate exposure into two margins: reallocation of demand across jobs and redesign of tasks within jobs. We document three main findings. First, generative AI exposure is dynamic rather than fixed, changing substantially over time. Second, labor demand adjusts through both margins. Hiring reallocation explains the largest share of the aggregate decline in exposure, accounting for 52\% on average, while within-job redesign becomes increasingly important, accounting for 39.5\%. A complementary Oaxaca–Blinder decomposition shows that shifts in occupational composition account for about 90\% of the exposure change attributable to observable job characteristics. Third, adjustment differs across the job ladder. Senior jobs adjust earlier and mainly through reallocation, whereas junior jobs adjust through a broader mix of reallocation, redesign, and their interaction. These findings suggest that labor-market adjustment to generative AI is a process of organizational reconfiguration, in which firms reshape both hiring demand and the task architecture of work.

    \vspace{1em}
    \end{abstract}

    \thispagestyle{empty}
\end{titlepage}

\setcounter{page}{1}
\doublespacing

\section{Introduction}
\label{sec:introduction}

Generative artificial intelligence (AI) has rapidly moved from a consumer-facing technology to a general-purpose tool used inside organizations. Since the public release of ChatGPT in late 2022, large language models have become increasingly embedded in writing, coding, data analysis, customer service, marketing, legal support, education, and managerial work \citep{hartley2026labor}. This diffusion has renewed a long-standing question in the social sciences: how does a major new technology reshape labor demand? The question has become especially urgent because generative AI differs from many earlier automation technologies. Earlier waves of automation were often associated with routine manual and rule-based work \citep{acemoglu2011skills, autor2015there, acemoglu2019automation}. Generative AI, in contrast, appears especially relevant for language-intensive, analytical, and creative tasks, i.e., activities that are central to many white-collar occupations.

The public debate has therefore focused heavily on displacement. Business leaders have warned that AI systems may reduce the need for some corporate roles.\footnote{Survey and industry evidence suggest that generative AI has begun to move from experimentation toward organizational deployment. McKinsey’s 2025 global survey reports that 23\% of respondents indicated that their organizations were scaling agentic AI in at least one business function, while another 39\% reported experimenting with AI agents \citep{singla2025state}. In parallel, senior executives at Anthropic, Amazon, and Ford have publicly linked AI diffusion to possible reductions in demand for some white-collar roles, especially as firms integrate generative AI and agentic AI into business functions \citep{cutter2025ceosjobs, reuters2025amazon, vandehei2025whitecollarbath}. Although such accounts should not be interpreted as systematic evidence, they motivate a broader empirical examination of how hiring demand and job design evolve during the diffusion of generative AI.} Media accounts have described fewer openings for entry-level white-collar workers \citep{iscenko2026looking}. Recent academic studies have asked whether employment or hiring has declined in occupations that appear more exposed to generative AI \citep{brynjolfsson2025canaries}. Yet the labor-market effects of a new technology need not appear only as fewer jobs in exposed occupations. Firms may also respond by changing the organization of work itself. They may shift hiring away from jobs whose baseline task content is more exposed to generative AI. They may also continue hiring within the same broad job categories while rewriting job descriptions, changing task requirements, and moving workers toward activities that are less substitutable by AI or more complementary to it. The first margin is a reallocation of hiring demand across jobs. The second is redesign of the tasks within jobs.

Distinguishing these two margins is central to understanding how generative AI affects work. If adjustment occurs only through hiring reallocation, the main empirical question is which occupations, industries, or seniority groups gain or lose labor demand. This is the perspective implicit in much of the emerging evidence linking occupation-level exposure to employment, productivity, wages, or job postings \citep{brynjolfsson2025generativeai, hampole2025artificial, chandar2025tracking, chen2025displacement, johnston2025labor, liu2025labor}. But if firms also redesign jobs, then the task content of work becomes endogenous to technological diffusion. In that case, exposure is not simply a fixed property of an occupation. It changes as employers revise what they ask workers to do. Labor-market adjustment then takes place even when broad occupational labels remain unchanged.

This distinction is also important for interpreting recent debates over the job ladder. Several studies and public reports have raised concerns that generative AI may be especially harmful to junior workers, whose tasks often involve drafting, summarizing, coding, analysis, and other activities that AI systems can assist or automate \citep{felten2023will, gmyrek2023generative, tomlinson2025working}. Other evidence questions this interpretation, arguing that declines in entry-level hiring may reflect broader macroeconomic forces rather than AI-driven substitution \citep{iscenko2026looking}. A key challenge is that many existing studies measure exposure at the occupation level and then compare employment or hiring across more- and less-exposed occupations \citep{eloundou2024gpts, brynjolfsson2025canaries}. Such approaches are useful for identifying where generative AI may matter most, but they cannot observe whether firms are changing the content of junior and senior jobs within the same occupation. As a result, they leave open a more general question: does generative AI merely change the composition of jobs firms post, or does it also change what those jobs contain?

This paper studies how generative AI reorganizes labor demand. We focus on three questions. First, how does exposure to generative AI in posted jobs evolve over time as the technology diffuses? Second, to what extent do changes in aggregate exposure reflect hiring reallocation across jobs versus redesign of task content within jobs? Third, do these adjustment margins differ across the job ladder?

Answering these questions requires data that observe labor demand at scale and contain information about the content of jobs. We use a nationwide dataset from Lightcast that contains all online job postings in the United States and covers all sectors of the economy from January 2021 through June 2025. The raw data contain more than 188 million postings during our study period. We implement a repeated random sampling procedure within occupation-by-seniority-by-industry cells, yielding a final sample of 9,373,092 postings. This nationwide coverage is important because the relevance of generative AI depends on the task content of work, which varies across occupations, industries, and seniority levels. Studies based on selected sectors or occupations may therefore capture only part of the labor-market adjustment. By covering the full U.S. economy, our data allow us to examine how changes across different parts of the labor market combine into aggregate patterns of labor-demand reorganization.

We construct a dynamic, posting-level measure of generative AI exposure using a two-stage large language model (LLM) pipeline. In the first stage, we extract tasks from each job description and match those tasks to skill groups (specialized or common). In the second stage, we classify each task according to whether current generative AI tools can substantially reduce the time required to complete it at equivalent quality. We then aggregate task-level labels into a posting-level exposure index. This design adapts the task-based logic of prior exposure measures to the level at which firms actually describe vacancies. It allows exposure to vary across postings within the same occupation, across industries and seniority levels, and over time as firms revise job content.

We then use two complementary decomposition methods to separate the margins of adjustment. The first is a three-fold extension of the Kitagawa decomposition \citep{kitagawa1955components}. We consider aggregate generative AI exposure as a weighted average of exposure across occupation-by-industry-by-seniority cells. Aggregate exposure can change because firms alter the mix of jobs they post, because the task content of similar jobs changes over time, or because both adjustments occur simultaneously. We interpret these components as hiring reallocation, job redesign, and their interaction. The second method is a weighted Oaxaca–Blinder decomposition comparing the pre- and post-GPT periods \citep{oaxaca1973male, oaxaca2025oaxaca}. This regression-based decomposition allows us to examine which observable job characteristics, especially occupation, industry, seniority, location, remote-work status, internship status, and employment type, account for the compositional change in exposure.

The analysis yields three main findings. First, generative AI exposure in posted jobs is dynamic. Mean exposure rises through early 2022, declines through 2023, and partially recovers thereafter. This pattern is difficult to reconcile with the view that exposure is a fixed attribute of occupations. It suggests instead that firms' stated task requirements evolve over time. The decline is especially concentrated among high-exposure occupations, while low- and medium-exposure occupations are more stable. Exposure also varies sharply across sectors. It is highest in finance and insurance, professional services, and information, and lowest in accommodation and food services, retail trade, and transportation and warehousing. Senior positions have higher exposure than junior or intermediate positions on average, revealing a seniority gradient that occupation-level measures cannot capture.

Second, labor demand adjusts through both hiring reallocation and job redesign. The composition effect turns negative after the third quarter of 2023, indicating that firms shift hiring away from jobs with higher baseline exposure. The within-cell exposure effect also becomes negative around the same period and grows in magnitude thereafter, indicating that firms reduce exposure within continuing job categories. From the third quarter of 2023 onward,\footnote{The timing is consistent with firms moving from experimentation with generative AI toward organizational integration, coinciding with the arrival of enterprise-grade tools such as
ChatGPT Enterprise \citep{openai2023chatgptenterprise}.} hiring reallocation accounts for 52\% of the decline in aggregate generative AI exposure, while within-cell job redesign accounts for 39.46\%. The interaction term accounts for the remaining 8.54\%. Thus, the decline in aggregate exposure is not simply a story of fewer postings in highly exposed occupations. A large share reflects changes in the task content of jobs that firms continue to post.

Third, adjustment differs across the job ladder. Senior jobs adjust earlier and primarily through hiring reallocation. From the third quarter of 2023 onward, the composition effect accounts for 70.80\% of the aggregate contribution among senior postings. Junior jobs show a broader pattern: hiring reallocation, job redesign, and their interaction all contribute meaningfully. Intermediate jobs track the aggregate pattern most closely, with reallocation and redesign contributing almost equally. These patterns suggest that generative AI does not simply move labor demand up or down the job ladder. It changes the margins through which firms adjust different layers of work. For senior roles, firms appear to adjust first by changing the structure of vacancies. For junior roles, firms simultaneously change both the types of jobs they post and the tasks embedded in those jobs.

The Oaxaca–Blinder decomposition reinforces these conclusions while clarifying the observable dimensions behind compositional adjustment. Average exposure declines after GPT diffusion, and both the explained and unexplained components are negative. The explained component accounts for roughly two-thirds of the aggregate decline, and occupational shifts account for about 90\% of the exposure decline attributable to observed job characteristics. Other posting-level characteristics, including remote-work arrangement, industry, employment type, and internship status, also contribute, although more modestly. These results show that occupational reallocation is the dominant observable source of compositional change, but it is not the whole story. The posting-level data reveal additional vacancy-design margins that are not visible in occupation-level analyses.

This paper contributes to research on technology and labor markets in three ways. First, it develops a dynamic, posting-level measure of generative AI exposure. Existing measures have been valuable for identifying which occupations are most exposed ex ante, but they generally assign a fixed score to occupations using standardized task taxonomies \citep{eloundou2024gpts, felten2023will, gmyrek2023generative, tomlinson2025working}. Our approach shows that exposure itself changes over time and varies within occupations. This matters because the diffusion of generative AI is not merely a shock to a fixed set of jobs; it is also a process through which firms update the task content of those jobs.

Second, the paper shifts the empirical focus from job displacement to labor-demand reorganization. Much of the current debate asks whether highly exposed occupations decline \citep{brynjolfsson2025canaries, chandar2025tracking, chen2025displacement, johnston2025labor, liu2025labor}. We ask how the structure of labor demand changes. This broader framing is important because firms can adapt to generative AI without eliminating an occupation or even reducing total hiring in a category. They can rewrite jobs, reweight tasks, and alter the mix of skills requested from workers. Our evidence suggests that such redesign is quantitatively large and becomes more important as generative AI tools become more organizationally deployable.

Third, the paper provides new evidence on generative AI and the job ladder. Prior work has debated whether junior workers are more exposed or more adversely affected than senior workers \citep{brynjolfsson2025canaries, hosseini2025generative}. We show that the relevant heterogeneity is not only about the magnitude of adjustment but also about its mechanism. Junior and senior jobs adjust through different combinations of hiring reallocation and task redesign. This distinction has implications for how new workers enter the labor market and acquire skills. If entry-level roles are being redesigned at the same time that firms are reallocating hiring away from exposed positions, then early-career workers may face not only fewer opportunities in some job categories but also a changing set of tasks within the opportunities that remain.

More broadly, the findings suggest that generative AI is reorganizing the architecture of work. The labor-market response is not captured by a simple substitution narrative in which exposed jobs decline and less-exposed jobs expand. Nor is it captured by a purely augmentation narrative in which the same jobs become more productive without changing their content. Instead, firms appear to adjust along multiple margins: they change where they hire, what jobs contain, and how these adjustments differ across hierarchy levels. For researchers, this implies that exposure measures should be treated as dynamic objects rather than fixed occupation-level characteristics. For firms, it suggests that workforce planning requires attention not only to headcount but also to task design. For policymakers and educators, it highlights the need to monitor how the entry points into professional work are changing as generative AI diffuses.

The remainder of the paper proceeds as follows. The next section reviews related research on AI exposure, labor demand, and the job ladder. Section 3 describes the Lightcast job-posting data and sampling procedure. Following that, Section 4 presents the construction of the posting-level generative AI exposure measure. Section 5 introduces the decomposition framework. Section 6 then reports the findings. Lastly, Section 7 concludes.

\section{Related Literature} 
\label{sec:literature}

This section situates our study in two related streams of research. The first develops measures of exposure to AI in general and, more specifically, generative AI. We build on this work by moving from static occupation-level exposure measures to a dynamic, posting-level measure that captures variation within occupations and over time. The second examines the labor-market effects of generative AI. We contribute to this literature by shifting attention from whether exposed jobs decline to how labor demand is reorganized through two margins: hiring reallocation and job redesign. We further examine how these margins differ across the job ladder.

\subsection{AI and Generative AI Exposure Metrics}

A large literature measures which types of work are likely to be affected by AI. The conceptual foundation of this literature is task-based: technology affects labor demand not by acting directly on occupations, but by changing the tasks performed within them \citep{acemoglu2011skills,autor2015there,acemoglu2019automation}. Because occupations are bundles of tasks, technological change may automate, augment, or otherwise reshape some parts of an occupation more than others. Consistent with this view, many AI exposure measures combine occupational task information from O*NET with external assessments of AI capabilities to construct occupation-level exposure indices \citep{frey2017future,felten2018method,felten2021occupational,webb2019impact,pizzinelli2023labor,hampole2025artificial}.

Recent work extends this approach to generative AI and LLMs. Much of this work continues to rely on O*NET task or ability information, but adapts the exposure concept to the capabilities of language models. \citet{eloundou2024gpts}, for example, develop a task-based rubric that evaluates whether LLMs and LLM-powered software can substantially reduce the time required to perform specific tasks, and then aggregate these assessments to the occupation level. \citet{felten2023will} adapt an ability-based occupational exposure framework to advances in language modeling. \citet{gmyrek2023generative} use task-level evaluations to distinguish automation from augmentation based on the distribution of exposure across tasks within occupations. \citet{benitez2024mirror} use synthetic AI surveys to construct a more holistic occupation-level measure based on occupations' characteristic tasks. Other studies use real-world interaction data. The Anthropic Economic Index \citep{handa2025economic} analyzes millions of Claude.ai conversations and maps them to O*NET tasks and occupations to identify where AI use is concentrated. \citet{tomlinson2025working} similarly use anonymized Microsoft Bing Copilot conversations, classify user goals and AI actions into O*NET work activities, and aggregate these classifications into occupation-level AI applicability scores.

These measures have been valuable for identifying which occupations are plausibly exposed to generative AI. However, most are constructed at the occupation level and are therefore fixed within occupations and largely time-invariant.\footnote{O*NET resurveys occupations on a rolling five-year cycle; between survey waves, the task profile for a given occupation remains unchanged. See \url{https://www.onetcenter.org/dataUpdates.html\#summary}.} This feature limits their ability to capture an important implication of the task-based view itself: if firms revise the tasks required in a job as technology diffuses, then exposure need not remain fixed even within the same occupation. Occupation-level measures can identify exposure ex ante, but they cannot observe whether employers subsequently rewrite job requirements, adjust skill demands, or change the task content of otherwise similar jobs.

We depart from this approach by constructing a dynamic, posting-level measure of generative AI exposure. Using job postings as the unit of analysis allows exposure to vary across postings within the same occupation and over time as employers revise task requirements. This granularity is central to our research design. It allows us to distinguish between two margins of labor-demand adjustment: hiring reallocation, in which firms change the mix of jobs they post, and job redesign, in which firms change the task content of comparable jobs. Static occupation-level measures assign a single exposure value to each occupation and therefore cannot separate these two channels.

\subsection{Effects of Generative AI in the Labor Market}

A rapidly growing literature examines whether generative AI has begun to affect labor-market outcomes. The emerging evidence is mixed, reflecting differences in data sources, exposure measures, outcomes, and empirical designs. We organize this literature around two issues most closely related to our analysis: changes in labor demand for exposed work and heterogeneity across the job ladder.

\textbf{Effects on employment and labor demand}: Several studies report evidence consistent with reduced demand for highly exposed or substitutable work. Using administrative payroll records, \citet{brynjolfsson2025canaries} document sizable relative employment declines in the most AI-exposed occupations after the release of ChatGPT. \citet{liu2025labor} find corresponding declines in job postings for more substitutable roles. Evidence from online labor markets points in a similar direction: studies generally find that generative AI reduces demand or earnings in highly automatable freelance tasks such as writing and translation \citep{qiao2023ai,hui2024short,demirci2025ai,teutloff2025winners}. Other work emphasizes heterogeneity in whether generative AI substitutes for or complements labor. \citet{chen2025displacement} show that occupations prone to automation experience declining labor demand and simplified skill requirements, whereas occupations prone to augmentation experience increases in demand and skill complexity. \citet{johnston2025labor} find average gains in wage bills and employment in more exposed sectors, alongside declines where AI is more directly substitutive.

At the same time, other studies question whether observed declines in exposed work can be attributed to generative AI. A central concern is that AI exposure is correlated with sensitivity to macroeconomic conditions. \citet{iscenko2026looking} show that highly AI-exposed occupations are concentrated in rate-sensitive sectors such as information, finance, and professional services, and that postings for these occupations began declining before the release of ChatGPT, around the onset of the Federal Reserve's sharp monetary tightening cycle. This suggests that part of the observed decline in demand for exposed work may reflect macroeconomic contraction rather than technological displacement. Related studies that attempt to detect employment effects directly also find limited evidence of aggregate labor-market effects. \citet{chandar2025tracking} find no systematic differences in employment patterns across more- and less-exposed occupations in Current Population Survey (CPS) data. \citet{humlum2026still} link survey-based chatbot adoption to Danish administrative records and report precisely estimated zero effects on individual earnings and hours worked, even as adopting workplaces experience task reorganization and create new AI-related roles.

This debate motivates our focus on mechanisms rather than only levels of employment or hiring. If changes in exposed labor demand partly reflect macroeconomic shocks, then aggregate declines in exposed occupations are difficult to interpret on their own. Our decomposition framework separates changes in aggregate exposure into shifts in the mix of posted jobs and changes in the task content of comparable jobs. The latter margin is especially informative because it captures variation within occupation-by-sector-by-seniority cells, rather than reallocation across these broad dimensions.

\textbf{Effects along the job ladder}: A related literature asks whether generative AI affects workers differently across career stages. \citet{brynjolfsson2025canaries} provide prominent evidence in this direction, documenting a 16 percent relative employment decline among early-career workers ages 22--25 in the most AI-exposed occupations after the release of ChatGPT. \citet{simon2025entrylevel} and \citet{eisfeldt2023generative} similarly document especially pronounced declines in postings for entry-level roles. Using firm-level data, \citet{hosseini2025generative} identify a seniority-biased pattern in which junior employment declines relative to senior employment within AI-adopting firms, driven mainly by slower junior hiring rather than increased separations.

The interpretation of these seniority-related patterns remains contested. \citet{iscenko2026looking} show that within highly AI-exposed occupations, postings for junior and senior roles have declined roughly in parallel since their peak in spring 2022, with little evidence that junior roles experienced disproportionately larger declines. They argue that some estimates of entry-level employment loss may reflect cohort dynamics during broad-based hiring slowdowns rather than seniority-targeted displacement. \citet{humlum2026still} reach a complementary conclusion in a difference-in-differences design, finding that AI adoption is not the main driver of observed declines in early-career employment in Denmark.

Our study contributes to this literature in three ways. First, we move beyond static occupation-level exposure measures by constructing a dynamic measure from the text of individual job postings. This allows us to observe how exposure changes within occupations as firms revise task requirements. Second, we distinguish two mechanisms of labor-demand adjustment. The Kitagawa decomposition separates changes in aggregate exposure into hiring reallocation across jobs and task redesign within comparable jobs. A complementary Oaxaca--Blinder decomposition then identifies which observable job characteristics are most associated with the exposure change. Third, we provide a more granular view of the job ladder. Existing studies often assign the same occupation-level exposure score to junior and senior jobs within an occupation. Our posting-level measure allows exposure to differ by seniority within the same occupation and allows us to examine whether junior and senior roles adjust through different margins. This distinction is important because generative AI may reshape career opportunities not only by changing demand for jobs at different seniority levels, but also by changing the task content through which workers enter, advance, and accumulate expertise.

\section{Data}
\label{sec:data}

\subsection{Lightcast Job Postings Data}

Our main data source is the U.S. online job postings database maintained by Lightcast.\footnote{Relevant documentation sources include Lightcast's job posting methodology: \url{https://kb.lightcast.io/en/articles/6957446-job-posting-analytics-jpa-methodology}, skills taxonomy: \url{https://kb.lightcast.io/en/articles/7216059-lightcast-skills-taxonomy}, and seniority definitions: \url{https://kb.lightcast.io/en/articles/11865067-job-seniority}.} Lightcast provides a large-scale, nationwide collection of vacancy advertisements posted by employers across industries, locations, and job types. It aggregates postings from more than 220{,}000 websites, including company career pages, national and local job boards, and job-posting aggregators. In the United States, Lightcast covers more than 435 million job postings since 2010. The data have been widely used in labor economics and related fields to study skill demand, technological change, and firm hiring behavior \citep{deming2018skill,acemoglu2022artificial,antoniades2025ai}.

These data are well suited to our research questions for two reasons. First, they allow us to observe posted labor demand at national scale. This is important because generative AI may affect different parts of the economy in different ways. A sample limited to selected occupations, industries, or firms may miss how adjustment in one segment of the labor market offsets, reinforces, or differs from adjustment elsewhere. Second, job postings contain rich textual descriptions of the tasks, skills, and responsibilities that employers associate with each vacancy. This allows us to measure generative AI exposure at the level of individual postings and to study how exposure changes as employers revise job content over time.

This combination of scale and textual detail distinguishes job postings data from other data sources commonly used in this literature. Occupation-level task databases such as O*NET provide standardized descriptions of work activities, but they assign a common task profile to an occupation and are not designed to capture rapid changes in job content within occupations.\footnote{See O*NET's Occupation Data documentation for more details at: \url{https://www.onetcenter.org/dictionary/30.2/excel/occupation_data.html}.} Employment or payroll data, such as those used in recent studies of AI and labor-market outcomes \citep{brynjolfsson2025canaries}, provide important evidence on realized employment but generally do not observe the task content of posted vacancies. Our setting requires both features: national coverage of labor demand and detailed information about what jobs contain. The Lightcast's job postings data allow us to examine two margins of adjustment to generative AI: reallocation in hiring demand across jobs and redesign of task content within comparable jobs.

The job description text is the primary input to our measurement strategy. We use it to extract posting-specific tasks and to construct each posting's exposure to generative AI. Unlike occupation-level exposure measures that assign the same score to all jobs within an occupation \citep{eloundou2024gpts}, posting text allows exposure to vary across jobs within the same occupation, across industries and seniority levels, and over time. This feature is central to our analysis because a main premise of the paper is that exposure is not fixed. It may evolve as firms change the tasks they request from their employees.

We complement the text with structured variables. Each posting is mapped to a standardized O*NET occupation code and a two-digit NAICS industry code, which allow us to compare jobs across occupations and sectors. We also use Lightcast's job seniority variable to classify postings into broad career stages. Lightcast identifies postings as Junior or Senior when the job title or posting text contains clear seniority language; postings without such language are classified as Intermediate. In addition, we use Lightcast's extracted skills data, including common and specialized skills, as inputs into our exposure-construction pipeline. These skill variables help organize posting text into skill groups, match extracted tasks to those groups, and weight tasks linked to specialized and common skills. Finally, we use other posting characteristics available in the data, including location, employment type, internship indicators, and remote-work indicators.

\subsection{Sampling Strategy}

Our sampling strategy is motivated by the scale of the raw data and the computational demands of our measurement approach. During our study period, from January 2021 through June 2025, the raw Lightcast data contain more than 188 million U.S. job postings. Because our exposure measure requires applying a two-stage large language model pipeline to job posting text, processing the full corpus is computationally infeasible. We therefore implement a repeated random sampling procedure designed to preserve the key sources of heterogeneity that are central to our research design: occupation, industry, seniority, and time.

We sample within cells defined by three dimensions. The first dimension is occupation, measured using the O*NET occupation code. Occupation is the natural starting point because much of the existing generative AI exposure literature measures exposure at the occupation or occupation-task level \citep{eloundou2024gpts,brynjolfsson2025canaries}. It is also likely to capture broad differences in the task content of work and in the potential applicability of generative AI.

The second dimension is industry, measured using the two-digit NAICS code. Even within the same occupation, jobs may involve different tasks across industries. For example, a data analyst, marketing specialist, or software developer may perform different activities depending on whether the employer is in finance, health care, manufacturing, retail, or professional services. Incorporating industry into the sampling design helps preserve this cross-sector heterogeneity and allows us to distinguish economy-wide labor-demand adjustment from changes concentrated in particular sectors.

The third dimension is seniority, measured using Lightcast's job seniority classification. Seniority is central to our analysis because recent research and public debate have raised the possibility that generative AI may affect junior and senior roles differently \citep{hampole2025artificial,brynjolfsson2025canaries}. Existing studies often assign the same occupation-level exposure score to all workers or postings within an occupation, which makes it difficult to observe whether exposure differs across career stages within the same occupation. By incorporating seniority directly into our sampling design, we ensure that junior, intermediate, and senior postings are represented within occupation-by-industry groups.

Operationally, for each half-year period from January 2021 through June 2025, we group postings into occupation $\times$ seniority $\times$ industry cells. This procedure yields 25{,}349 cells in total. We drop cells with fewer than 20 postings in a given half-year period. These sparse cells account for approximately 0.45\% of all postings, so the restriction removes only a negligible share of the raw data while reducing noise from very small cells. We then draw a 5\% random sample from the remaining postings within each occupation $\times$ seniority $\times$ industry cell-period. This repeated cell-period sampling procedure yields a final sample of 9{,}373{,}092 postings.

The sampling design serves two purposes. First, it makes the LLM-based measurement task computationally feasible while retaining a large, nationwide sample of job postings. Second, it preserves the empirical variation needed for our decomposition analyses. Because the sample is drawn within occupation-by-industry-by-seniority cells over time, it maintains the structure required to study both changes in the composition of posted labor demand and changes in exposure within comparable jobs.

\begin{figure}[htbp]
    \centering
    \includegraphics[width=1\textwidth]{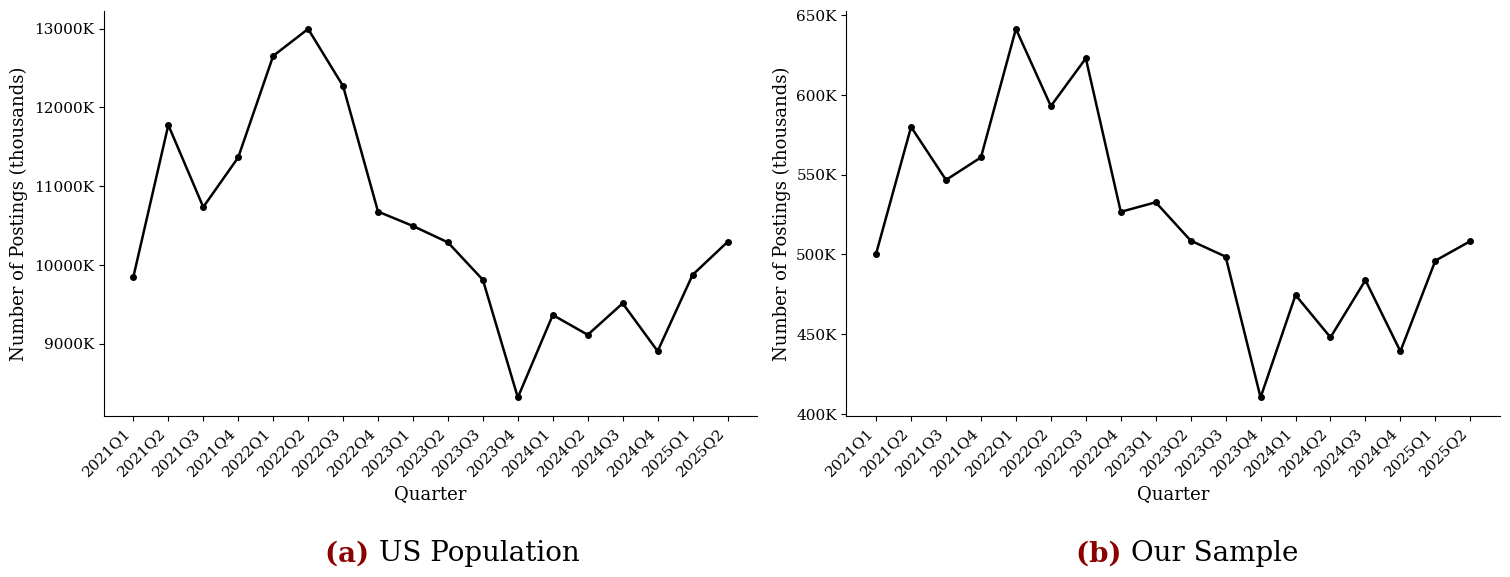}
    \caption{Quarterly Number of Job Postings in the U.S. Population and Our Sample}
    \label{fig:sample_population_quarterly}
    \vspace{0.3cm}

    \begin{minipage}{1\textwidth}
        \footnotesize
        \textit{Notes}: This figure compares the quarterly number of postings in the full data (Panel (a)) and in our sampled data (Panel (b)) from Quarter 1, 2021 to Quarter 2, 2025.
    \end{minipage}
    \vspace{-9pt}
\end{figure}

Figure~\ref{fig:sample_population_quarterly} compares the quarterly number of postings in the full nationwide Lightcast data and in our sampled data. The two series track each other closely. Both rise through 2021 and early 2022, peak around the second quarter of 2022, decline throughout 2023, and begin to recover in early 2024. The close alignment indicates that our sampling procedure preserves the main aggregate dynamics of the underlying population.


\section{Measurement of Posting-Level Exposure to Generative AI}
\label{sec:measurement}

We measure exposure to generative AI at the level of individual job postings. This measurement strategy is central to our research design. Most existing exposure measures assign a fixed score to an occupation based on standardized task descriptions. Such measures are useful for identifying which occupations are plausibly exposed to generative AI, but they cannot observe how exposure varies across postings within the same occupation or how exposure changes as employers revise job content. We instead use the text of each job posting to recover the tasks employers describe and to classify the exposure of those tasks to generative AI.

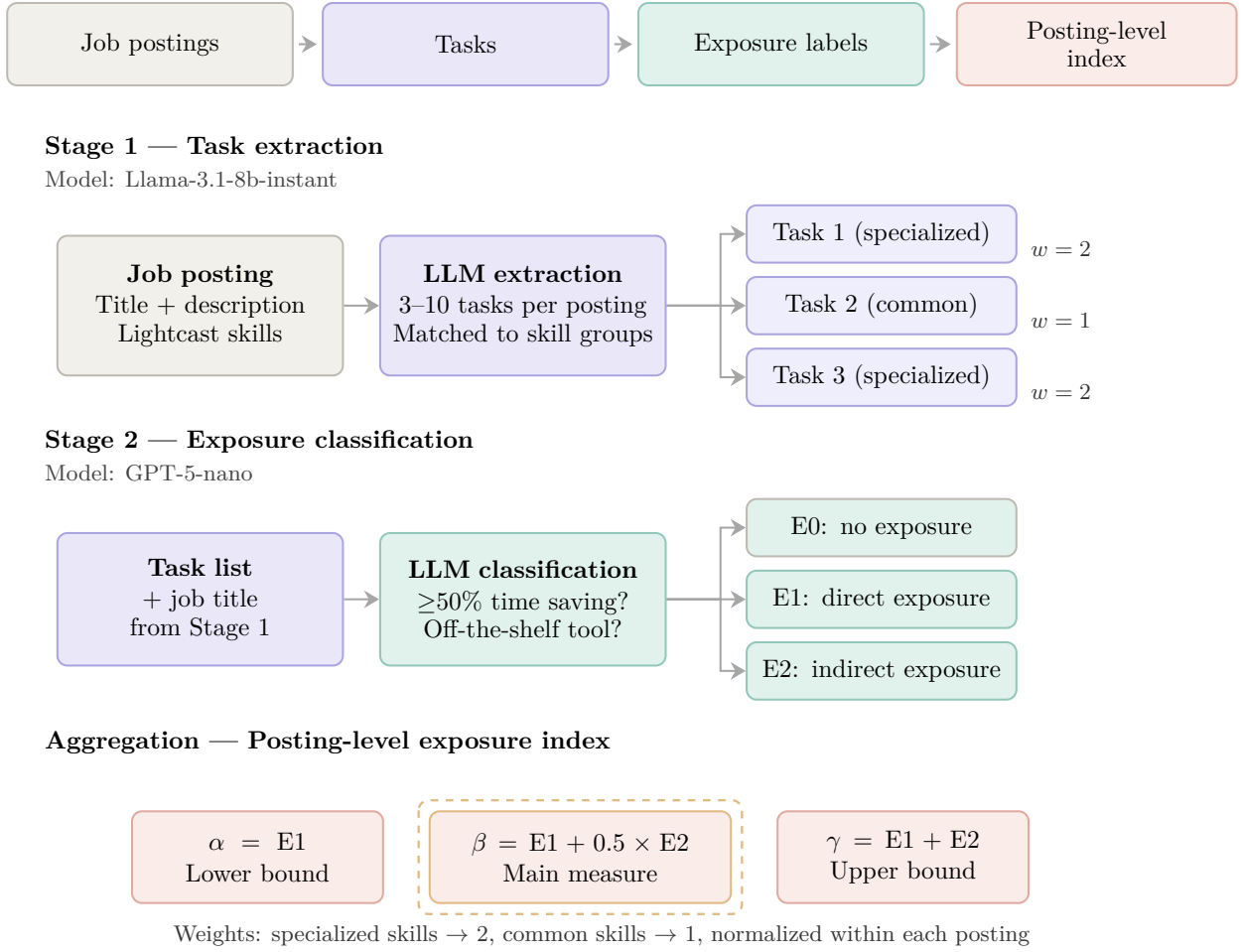
\begin{figure}[h]
  \centering
  \resizebox{1\textwidth}{!}{\begin{tikzpicture}[x=1cm,y=1cm]

\node[mybox, fill=boxgray, draw=boxgrayline, minimum height=1.15cm] 
    (jp) at (-6.0,0) {Job postings};

\node[mybox, fill=boxpurple, draw=boxpurpleline, minimum height=1.15cm] 
    (tk) at (-1.6,0) {Tasks};

\node[mybox, fill=boxgreen, draw=boxgreenline, minimum height=1.15cm] 
    (el) at (2.8,0) {Exposure labels};

\node[mybox, fill=boxpink, draw=boxpinkline, minimum height=1.15cm, text width=3.6cm] 
    (pi) at (7.2,0) {Posting-level\\index};

\draw[arrow, shorten <=2pt, shorten >=2pt] (jp.east) -- (tk.west);
\draw[arrow, shorten <=2pt, shorten >=2pt] (tk.east) -- (el.west);
\draw[arrow, shorten <=2pt, shorten >=2pt] (el.east) -- (pi.west);

\node[stageLabel] at (-7.6,-1.45) {Stage 1 --- Task extraction};
\node[modelLabel] at (-7.6,-1.88) {Model: Llama-3.1-8b-instant};

\node[mybox, fill=boxgray, draw=boxgrayline, minimum height=1.95cm] 
    (s1a) at (-5.3,-3.65)
    {\textbf{Job posting}\\[1pt]
     Title + description\\
     Lightcast skills};

\node[mybox, fill=boxpurple, draw=boxpurpleline, minimum height=1.95cm]
    (s1b) at (-0.8,-3.65)
    {\textbf{LLM extraction}\\[1pt]
     3--10 tasks per posting \\
     Matched to skill groups};

\node[slimbox, fill=taskblue, draw=taskblueline] (t1) at (4.2,-2.65) {Task 1 (specialized)};
\node[slimbox, fill=taskblue, draw=taskblueline] (t2) at (4.2,-3.65) {Task 2 (common)};
\node[slimbox, fill=taskblue, draw=taskblueline] (t3) at (4.2,-4.65) {Task 3 (specialized)};

\node[noteFont, anchor=west] at (6.15,-2.85) {$w = 2$};
\node[noteFont, anchor=west] at (6.15,-3.85) {$w = 1$};
\node[noteFont, anchor=west] at (6.15,-4.85) {$w = 2$};

\draw[arrow] (s1a.east) -- (s1b.west);
\draw[arrow] (s1b.east) -- ($(s1b.east)+(0.75,0)$) |- (t1.west);
\draw[arrow] (s1b.east) -- ($(s1b.east)+(0.75,0)$) |- (t2.west);
\draw[arrow] (s1b.east) -- ($(s1b.east)+(0.75,0)$) |- (t3.west);

\node[stageLabel] at (-7.6,-5.55) {Stage 2 --- Exposure classification};
\node[modelLabel] at (-7.6,-5.98) {Model: GPT-5-nano};

\node[mybox, fill=boxpurple, draw=boxpurpleline, minimum height=1.85cm]
    (s2a) at (-5.3,-7.75)
    {\textbf{Task list}\\[1pt]
     + job title\\
     from Stage 1};

\node[mybox, fill=boxgreen, draw=boxgreenline, minimum height=1.85cm]
    (s2b) at (-0.8,-7.75)
    {\textbf{LLM classification}\\[1pt]
     $\geq$50\% time saving?\\
     Off-the-shelf tool?};

\node[slimbox, fill=boxgreen, draw=boxgrayline]     (e0) at (4.2,-6.75) {E0: no exposure};
\node[slimbox, fill=boxgreen, draw=boxgreenline]   (e1) at (4.2,-7.75) {E1: direct exposure};
\node[slimbox, fill=boxgreen, draw=boxgreenline] (e2) at (4.2,-8.75) {E2: indirect exposure};

\draw[arrow] (s2a.east) -- (s2b.west);
\draw[arrow] (s2b.east) -- ($(s2b.east)+(0.75,0)$) |- (e0.west);
\draw[arrow] (s2b.east) -- ($(s2b.east)+(0.75,0)$) |- (e1.west);
\draw[arrow] (s2b.east) -- ($(s2b.east)+(0.75,0)$) |- (e2.west);

\node[stageLabel] at (-7.6,-9.75) {Aggregation --- Posting-level exposure index};

\node[mybox, fill=boxpink, draw=boxpinkline, minimum height=1.28cm, minimum width=3.5cm, text width=3.2cm]
    (a) at (-4.5,-11.35)
    {$\alpha = \mathrm{E1}$\\[1pt]Lower bound};

\node[mybox, fill=boxpink, draw=expyellowline, minimum height=1.28cm, minimum width=4.2cm, text width=3.9cm]
    (b) at (0,-11.35)
    {$\beta = \mathrm{E1} + 0.5 \times \mathrm{E2}$\\[1pt]Main measure};

\node[mybox, fill=boxpink, draw=boxpinkline, minimum height=1.28cm, minimum width=3.5cm, text width=3.2cm]
    (g) at (4.5,-11.35)
    {$\gamma = \mathrm{E1} + \mathrm{E2}$\\[1pt]Upper bound};

\draw[dashed, line width=0.9pt, draw=expyellowline, rounded corners=4pt]
  ($(b.north west)+(-0.14,0.14)$) rectangle ($(b.south east)+(0.14,-0.14)$);

\node[noteFont, align=center] at (0.3,-12.45)
{Weights: specialized skills $\rightarrow$ 2, common skills $\rightarrow$ 1, normalized within each posting};

\end{tikzpicture}}
  \caption{Two-Stage LLM Pipeline for Computing Posting-Level AI Exposure Indices}
  \label{fig:exposure-pipeline}
  \vspace{-9pt}
\end{figure}

Our approach follows the task-based logic of \citet{eloundou2024gpts}, but adapts it from occupation-level task data to posting-level job descriptions. The resulting measure allows exposure to vary across occupations, industries, seniority levels, and time. It also allows us to distinguish the two adjustment margins. If firms change the mix of postings across jobs with different baseline exposure, aggregate exposure changes through hiring reallocation. If firms change the task content of comparable jobs, aggregate exposure changes through job redesign. A posting-level exposure measure is necessary to observe the latter margin. Figure~\ref{fig:exposure-pipeline} summarizes the two-stage large language model (LLM) pipeline used to compute posting-level exposure.


\subsection{Two-Stage LLM Pipeline}

The pipeline proceeds in two stages. In the first stage, we use \texttt{Llama-3.1-8b-instant} to extract tasks from each job posting and to link each task to a skill group constructed from the extracted skill information by Lightcast. In the second stage, we use \texttt{GPT-5-nano} to classify each extracted task into an exposure tier based on the extent to which currently available generative AI tools can perform or assist the task. The full prompts used in both stages are reported in ~\ref{app:full_prompts}.

\noindent\textbf{Stage 1: Task extraction from job descriptions.}
\label{stage1}

The first stage transforms unstructured posting text into a structured set of posting-specific tasks. The LLM receives the job title, job description, and the specialized and common skills extracted by Lightcast. It is instructed to identify the concrete work activities described in the posting rather than infer generic tasks from the occupation title. This distinction is important because two postings in the same occupation may emphasize different responsibilities, tools, deliverables, or work contexts.

For each posting, the model extracts between 3 and 10 tasks. It also uses the posting's Lightcast skills to form semantically related skill groups while preserving the distinction between specialized and common skills. Each extracted task is matched to exactly one skill group. This match determines whether the task is associated with a specialized skill or a common skill. We use this distinction to assign task-importance weights. Tasks matched to specialized skills receive a raw weight of 2, while tasks matched to common skills receive a raw weight of 1.\footnote{The specialized/common distinction is assigned through the task--skill-group match rather than through a direct task classification. The Stage 1 prompt groups Lightcast-provided specialized and common skills separately and assigns each extracted task to the closest skill group, as shown in ~\ref{app:full_prompts}. Tasks matched to specialized-skill groups are treated as specialized-skill tasks; tasks matched to common-skill groups are treated as common-skill tasks. In ties, the prompt gives priority to specialized-skill groups.} The purpose of this weighting is to give more influence to tasks linked to occupation- or role-specific skill requirements, while still retaining tasks associated with general workplace skills. 

\noindent\textbf{Stage 2: Exposure classification.}
\label{stage2}

The second stage classifies the exposure of each extracted task to generative AI. The model receives the job title and the Stage-1 task list and assigns each task exactly one exposure label from the set \(\{E0,E1,E2\}\). The labels are designed to capture whether currently available generative AI tools can substantially reduce the time required to complete the task while maintaining equivalent quality. Following \citet{eloundou2024gpts}, we define a task as exposed if generative AI can reduce completion time by at least 50\% at equivalent quality. Equivalent quality means that a third party receiving or evaluating the output would not notice or care that generative AI assistance was used.

The three labels distinguish different degrees of exposure. \(E0\) denotes tasks for which current off-the-shelf generative AI tools are unlikely to generate a meaningful productivity improvement, or for which using such tools would materially reduce output quality. \(E1\) denotes tasks that can be directly assisted by a single off-the-shelf generative AI or LLM tool without special integration, fine-tuning, or workflow redesign. \(E2\) denotes tasks for which a standalone tool may not be sufficient, but for which a thin AI-powered software layer or workflow integration could plausibly generate a substantial productivity improvement. Table~\ref{tab:exposure_rubric} summarizes the rubric.

\subsection{From Task-Level Labels to Posting-Level Exposure}
\label{subsec:task_to_posting}

After the two-stage annotation is complete, we aggregate task-level exposure labels into posting-level exposure measures. The objective is to summarize, for each vacancy, the extent to which the tasks described in the posting are exposed to generative AI. This posting-level aggregation is important because our analysis treats the job posting, rather than the occupation, as the basic unit at which employers describe labor demand.

Each extracted task is matched in Stage 1 to either a specialized-skill group or a common-skill group based on the posting's Lightcast skills. We use this distinction to assign task-importance weights. Following the logic in \citet{eloundou2024gpts}, who assign greater weight to core tasks than to supplemental tasks at the occupation level, we assign greater weight to tasks linked to specialized skills. Specifically, tasks matched to specialized skills receive a raw weight of 2, while tasks matched to common skills receive a raw weight of 1. This weighting scheme reflects the idea that specialized-skill tasks are more central to the role-specific content of a posting, whereas common-skill tasks capture more general workplace activities. Table~\ref{tab:exposure_examples} illustrates this mapping for two postings and shows how task-level exposure labels are aggregated to the posting level.


\begin{table}[t]
\vspace{3pt}
\caption{Summary of Exposure Rubric}
\vspace{-9pt}
\label{tab:exposure_rubric}
\begin{minipage}{0.95\textwidth}
\setlength{\fboxsep}{16pt}
\fbox{
\begin{minipage}{0.96\textwidth}

\textit{No exposure (\(E0\)) if:}
\vspace{-9pt}
\begin{itemize}
    \item a single off-the-shelf generative AI or LLM tool cannot reduce the time required to complete the task by at least 50\% while maintaining equivalent quality; or using such tools would materially reduce output quality.
\end{itemize}

\textit{Direct exposure (\(E1\)) if:}
\vspace{-9pt}
\begin{itemize}
    \item a single off-the-shelf generative AI or LLM tool, with no special integrations or fine-tuning, can reduce the time required to complete the task by at least 50\% at equivalent quality.
\end{itemize}

\textit{Indirect exposure (\(E2\)) if:}
\vspace{-9pt}
\begin{itemize}
    \item a single off-the-shelf generative AI or LLM tool alone cannot reduce the time required to complete the task by at least 50\%; but
    \item a thin AI-powered software layer built on top of such a tool could plausibly achieve at least a 50\% reduction while maintaining equivalent quality.
\end{itemize}

\end{minipage}
}
\end{minipage}
\vspace{-6pt}
\end{table}
\FloatBarrier

Formally, let posting \(p\) contain tasks \(j=1,\dots,J_p\). Let \(raw_{p,j}\) denote the raw weight assigned to task \(j\):
\setlength{\abovedisplayskip}{4pt}
\setlength{\belowdisplayskip}{4pt}
\begin{equation*}
raw_{p,j} =
\begin{cases}
2 & \text{if task } j \text{ is matched to a specialized-skill group,} \\
1 & \text{if task } j \text{ is matched to a common-skill group.}
\end{cases}
\end{equation*}

We normalize these raw weights within each posting:
\begin{equation*}
w_{p,j} = \frac{raw_{p,j}}{\sum_{k=1}^{J_p} raw_{p,k}},
\qquad \text{so that } \sum_{j=1}^{J_p} w_{p,j} = 1.
\end{equation*}


\begin{table}[htbp]
\centering
\caption{Illustrative Examples of Task-Level Exposure Classification}
\vspace{-9pt}
\label{tab:exposure_examples}
\begin{threeparttable}
\setlength{\tabcolsep}{9.5pt}

\footnotesize
\renewcommand{\arraystretch}{1.3}
\begin{tabular}{p{0.7cm} p{8.2cm} c c c}
\toprule
\textbf{ID} & \textbf{Task} & \textbf{Skill type} & \textbf{Weight} & \textbf{Exposure} \\
\midrule

\multicolumn{5}{l}{\textbf{Panel A: Technical Consultants}} \\
\multicolumn{5}{p{14cm}}{
O*NET 15-1299.00, Senior, Professional Services, 2025-02-27.
\newline
\(E1=0.73\), \(E2=0.27\), \(E0=0.00\), \(\beta=0.87\).
} \\
\addlinespace 
t1 & Design, develop, and maintain web applications using Node.js, React, and TypeScript. & Specialized & 2 & E1 \\
t2 & Develop custom AI agents for searching documents and information stored in SharePoint Online, Microsoft Teams, OneDrive, and other enterprise document storage systems. & Specialized & 2 & E2 \\
t3 & Build and support applications using SharePoint Framework (SPFx) solutions and Power Platform. & Specialized & 2 & E1 \\
t4 & Develop applications using Power Platform and build integrations for accessing business data in AWS data lake, SAP, and cloud database services. & Specialized & 2 & E1 \\
t5 & Design and develop custom SharePoint (SPFx) solutions based on business requirements. & Specialized & 2 & E1 \\
t6 & Stay updated with the latest trends, tools, and best practices in web development and software development. & Common & 1 & E1 \\
t7 & Contribute to the continuous improvement of the development process, tools, and methodologies. & Specialized & 2 & E2 \\
t8 & Deploy business apps on AWS and Azure cloud platforms. & Specialized & 2 & E1 \\

\midrule
\multicolumn{5}{l}{\textbf{Panel B: Retail Sales Associate}} \\
\multicolumn{5}{p{14cm}}{
O*NET 41-2031.00, Junior, Retail Trade, 2025-02-01.
\newline
\(E1=0.00\), \(E2=0.14\), \(E0=0.86\), \(\beta=0.07\).
} \\
\addlinespace

t1 & Greet customers and assist them in locating merchandise in the store. & Common & 1 & E0 \\
t2 & Explain product features and answer customer questions. & Common & 1 & E0 \\
t3 & Process sales transactions using point-of-sale (POS) systems. & Specialized & 2 & E0 \\
t4 & Handle cash, credit, and digital payments accurately. & Specialized & 2 & E0 \\
t5 & Maintain store appearance by stocking shelves and organizing displays. & Specialized & 2 & E0 \\
t6 & Monitor inventory levels and report stock shortages. & Specialized & 2 & E2 \\
t7 & Resolve customer complaints and provide after-sales support. & Common & 1 & E0 \\
\bottomrule
\end{tabular}

\begin{tablenotes}
\footnotesize
\item \textit{Notes}: The table reports two illustrative postings. Specialized-skill tasks receive a raw weight of 2, and common-skill tasks receive a raw weight of 1. Exposure labels are assigned at the task level. The posting-level measure is computed as \(\beta = E1 + 0.5 \times E2\), where \(E0\), \(E1\), and \(E2\) are weighted task shares.
\end{tablenotes}
\end{threeparttable}
\end{table}

Let \(I_{p,j}^{(k)}\) be an indicator equal to one if task \(j\) in posting \(p\) is classified into exposure tier \(E_k\), where \(k \in \{0,1,2\}\). We define the weighted share of posting \(p\)'s task content in exposure tier \(E_k\) as
\begin{equation}
\label{eq:weighted_share}
shareE_{k,p} = \sum_{j=1}^{J_p} w_{p,j} I_{p,j}^{(k)}.
\end{equation}

Thus, \(shareE_{0,p}\), \(shareE_{1,p}\), and \(shareE_{2,p}\) denote the weighted shares of task content with no exposure, direct exposure, and indirect exposure, respectively. Because each task is assigned to exactly one exposure tier and task weights sum to one, these shares satisfy
\begin{equation*}
shareE_{0,p} + shareE_{1,p} + shareE_{2,p} = 1.
\end{equation*}

Using these shares, we construct three posting-level exposure indices:
\begin{align}
\alpha_p &= shareE_{1,p}, \label{eq:alpha} \\
\beta_p  &= shareE_{1,p} + 0.5\,shareE_{2,p}, \label{eq:beta} \\
\gamma_p &= shareE_{1,p} + shareE_{2,p}. \label{eq:gamma}
\end{align}

The first measure, \(\alpha_p\), counts only directly exposed tasks and can be interpreted as a conservative lower-bound measure of exposure. The second measure, \(\beta_p\), fully counts directly exposed tasks and assigns a weight of 0.5 to indirectly exposed tasks. This is our main measure. The third measure, \(\gamma_p\), fully counts both directly and indirectly exposed tasks and can be interpreted as an upper-bound measure.

We use \(\beta_p\) as the primary exposure index because it captures the distinction between immediate and more implementation-dependent exposure. Tasks classified as \(E1\) can be assisted by off-the-shelf generative AI tools with little additional organizational change. Tasks classified as \(E2\) are also exposed, but their exposure is less immediate because meaningful productivity gains likely require complementary software, workflow integration, or organizational adoption. Assigning \(E2\) a partial weight therefore reflects the idea that indirect exposure is economically meaningful but less direct than \(E1\). This approach is consistent with prior work that distinguishes direct and indirect generative AI exposure \citep{eloundou2024gpts,brynjolfsson2025canaries,chen2025displacement}.

The resulting index \(\beta_p\) ranges from 0 to 1. A value of 0 indicates that all weighted tasks in a posting are classified as unexposed. A value of 1 indicates that all weighted tasks are directly exposed. Intermediate values capture the weighted mix of unexposed, directly exposed, and indirectly exposed tasks in the posting. Because \(\beta_p\) is constructed from posting-specific tasks, it can vary across postings within the same occupation and over time. This feature is essential for our decomposition analysis: changes in aggregate exposure may reflect both shifts in the distribution of postings across jobs and changes in the task composition of comparable postings.

~\ref{app:add_des_figs} shows that the time-series patterns of \(\alpha_p\), \(\beta_p\), and \(\gamma_p\) are similar. This suggests that our main conclusions are not sensitive to the particular weighting of indirectly exposed tasks.

\subsection{Descriptive Patterns}
\label{subsec:descriptive_patterns}

We begin by documenting several descriptive patterns in the posting-level exposure measure. These patterns serve two purposes. First, they show that generative AI exposure varies substantially across postings, seniority levels, occupations, industries, and time. Second, they motivate the decomposition analyses that follow by showing why aggregate changes in exposure may reflect both changes in the mix of posted jobs and changes in the task content of comparable jobs.

Table~\ref{tab:summary_exposure} reports summary statistics for the posting-level exposure measures overall and by job seniority. In the full sample, 48.9\% of weighted task content is classified as unexposed (\(E0\)), 28.2\% as directly exposed (\(E1\)), and 22.9\% as indirectly exposed (\(E2\)). These task shares imply mean posting-level exposure values of 0.282 for \(\alpha\), 0.396 for \(\beta\), and 0.511 for \(\gamma\).


\begin{table}[htbp]
\centering
\caption{Summary Statistics of Posting-Level Generative AI Exposure}
\vspace{-9pt}
\footnotesize
\label{tab:summary_exposure}
\begin{threeparttable}
\setlength{\tabcolsep}{12pt}

\begin{tabular}{lcccccccc}
\toprule
& \multicolumn{2}{c}{All} & \multicolumn{2}{c}{Junior} & \multicolumn{2}{c}{Intermediate} & \multicolumn{2}{c}{Senior} \\
\cmidrule(lr){2-3}\cmidrule(lr){4-5}\cmidrule(lr){6-7}\cmidrule(lr){8-9}
& Mean & SD & Mean & SD & Mean & SD & Mean & SD \\
\midrule
\multicolumn{9}{l}{\textit{Panel A. Exposure Composition}} \\
Share in $E0$ & 0.489 & 0.375 & 0.453 & 0.360 & 0.511 & 0.374 & 0.218 & 0.285 \\
Share in $E1$ & 0.282 & 0.263 & 0.306 & 0.261 & 0.273 & 0.261 & 0.383 & 0.268 \\
Share in $E2$ & 0.229 & 0.266 & 0.241 & 0.258 & 0.216 & 0.260 & 0.399 & 0.294 \\
\addlinespace
\multicolumn{9}{l}{\textit{Panel B. Exposure Measures}} \\
$\alpha$ & 0.282 & 0.263 & 0.306 & 0.261 & 0.273 & 0.261 & 0.383 & 0.268 \\
$\beta$  & 0.396 & 0.296 & 0.426 & 0.287 & 0.381 & 0.296 & 0.582 & 0.234 \\
$\gamma$ & 0.511 & 0.375 & 0.547 & 0.360 & 0.489 & 0.374 & 0.782 & 0.285 \\
\addlinespace
Number of postings & \multicolumn{2}{c}{9,373,092} & \multicolumn{2}{c}{655,229} & \multicolumn{2}{c}{8,135,089} & \multicolumn{2}{c}{582,774} \\
\bottomrule
\end{tabular}

\begin{tablenotes}[flushleft]
\scriptsize
\item \textit{Notes}: $E0$, $E1$, and $E2$ denote the weighted shares of posting-level task content classified into no exposure, direct exposure, and indirect exposure, respectively. The three exposure measures are defined as $\alpha = E1$, $\beta = E1 + 0.5 \times E2$, and $\gamma = E1 + E2$.
\end{tablenotes}
\end{threeparttable}
\vspace{-3pt}
\end{table}

Exposure differs systematically across the job ladder. Senior postings have the highest exposure across all measures, with a mean \(\beta\) of 0.582. Junior postings follow, with a mean \(\beta\) of 0.426, while intermediate postings have a mean \(\beta\) of 0.381. The seniority gradient is visible in both direct and indirect exposure: senior postings contain lower shares of unexposed task content and substantially higher shares of both \(E1\) and \(E2\) tasks. This pattern suggests that the exposure of work to generative AI is not only an occupation-level phenomenon. It also varies across career stages within the labor market.

This evidence highlights a central advantage of our posting-level measurement strategy. Existing generative AI exposure measures are typically constructed at the occupation level using standardized task taxonomies such as O*NET, assigning a common exposure score to jobs within an occupation regardless of seniority \citep{eloundou2024gpts,felten2023will}. Our pipeline recovers an additional source of heterogeneity: postings at different seniority levels can describe systematically different tasks even when they belong to the same broad occupation. This heterogeneity is central to our analysis of the job ladder.

As an external benchmark, we replicate the occupation-level exposure measure in \citet{eloundou2024gpts} and compare it with our posting-level measure. ~\ref{app:compare_eloundou} reports the comparison. When aggregated to the occupation level, our posting-based measure is highly correlated with the O*NET-based occupation-level benchmark. This provides evidence that our pipeline captures broad occupation-level exposure patterns identified in prior work. At the same time, our measure captures additional variation within occupations, across seniority levels, and over time. These dimensions of heterogeneity are absent from a single time-invariant occupation-level score.

\begin{figure}[htbp]
    \centering
    \includegraphics[width=1\textwidth]{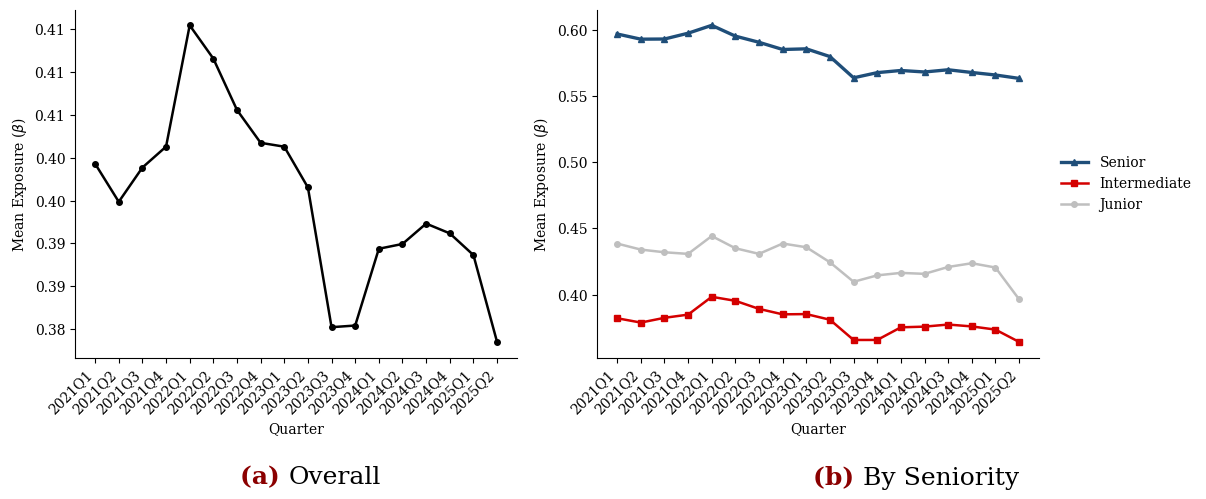}
    \caption{Quarterly Trend in Mean Generative AI Exposure ($\beta$)}
    \label{fig:beta_trend_main}
    \vspace{0.2cm}

    \begin{minipage}{1\textwidth}
        \footnotesize
        \textit{Notes}: This figure plots the quarterly trend in our main exposure measure, $\beta$, from Quarter 1, 2021 to Quarter 2, 2025. Panel (a) shows the overall mean in the sample. Panel (b) reports the same series separately for junior, intermediate, and senior postings.
    \end{minipage}
    \vspace{-6pt}
\end{figure}

\textbf{Exposure across time and seniority}: Figure~\ref{fig:beta_trend_main} plots the quarterly trend in our main exposure measure, \(\beta\). Panel (a) shows the overall trend. Mean exposure rises through early 2022, reaching a peak of 0.415, then declines through 2023 to a trough of 0.378 before partially recovering later in the sample. This pattern shows that exposure is not fixed over time. Because the measure is constructed from posting-specific tasks, the time-series movement may reflect changes in the distribution of postings across jobs, changes in task content within comparable jobs, or both.


Panel (b) plots the same series separately by seniority. Senior postings have consistently higher exposure than junior and intermediate postings throughout the sample period. Junior postings also exhibit higher exposure than intermediate postings. These level differences are consistent with the summary statistics in Table~\ref{tab:summary_exposure} and motivate our subsequent analysis of whether generative AI adjustment differs across the job ladder. ~\ref{app:add_des_figs} reports corresponding trends for \(E0\), \(E1\), \(E2\), and the alternative exposure measures \(\alpha\) and \(\gamma\).


\begin{figure}[htbp]
    \centering
    \includegraphics[width=0.9\textwidth]{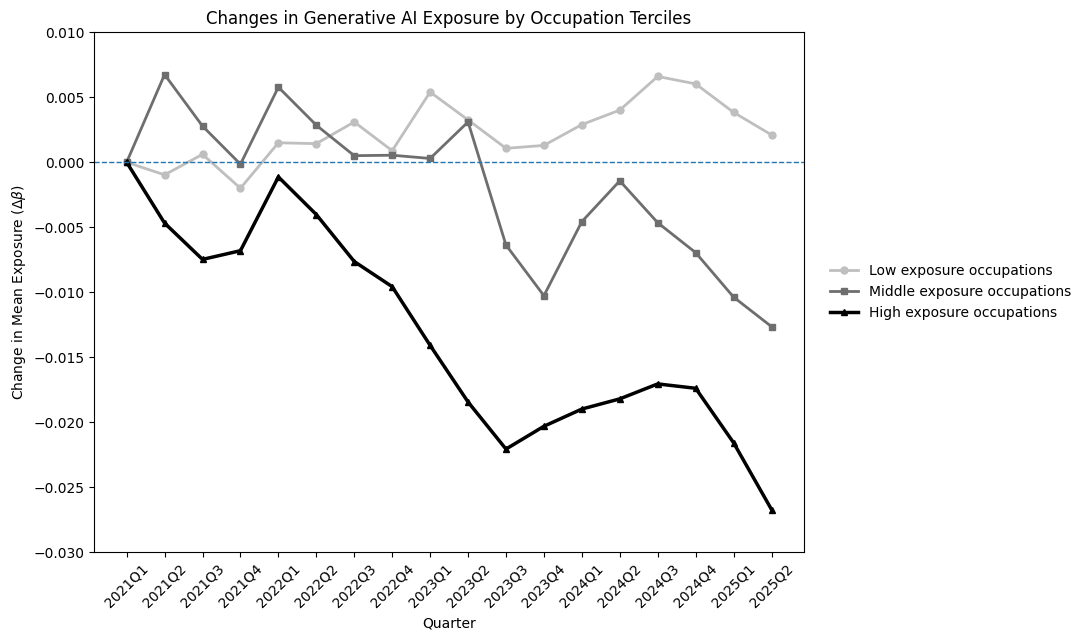}
    \caption{Changes in Generative AI Exposure by Occupation Group}
    \label{fig:occupation_terciles_beta}
    \vspace{0.2cm}

    \begin{minipage}{1\textwidth}
        \footnotesize
        \textit{Notes}: This figure groups occupations into three categories based on their average posting-level exposure, $\beta$, over the full sample period: low-, middle-, and high-exposure occupations. It then plots the quarterly change in mean exposure for each group. The dashed horizontal line indicates zero change.
    \end{minipage}
\end{figure}

\textbf{Exposure across occupations}: Figure~\ref{fig:occupation_terciles_beta} provides an occupation-level view of exposure dynamics. We divide occupations into terciles based on their average \(\beta\) over the full sample period and plot quarterly changes in mean exposure for low-, medium-, and high-exposure occupations. The figure shows that exposure dynamics differ substantially across occupation groups. The largest declines occur in the high-exposure group, while the low-exposure group changes relatively little. This pattern suggests that the aggregate decline in exposure is not uniform across the market; it is concentrated among occupations whose task content was initially more exposed to generative AI.

~\ref{app:add_des_figs} further splits these occupation groups by seniority. The broad tercile pattern persists across seniority levels, especially in the high-exposure group, where junior, intermediate, and senior postings all exhibit sustained declines over time. Within high-exposure occupations, junior postings show the largest decline in exposure, suggesting that task-content adjustment may be especially pronounced for junior roles in occupations that were initially more exposed. ~\ref{app:top_bottom_occ} reports the 20 occupations with the highest and lowest average exposure. The most exposed occupations are concentrated in writing, content creation, digital, and analytical work, whereas the least exposed occupations involve manual, physical, or routine operational tasks. This pattern is consistent with prior evidence that generative AI is especially relevant for language- and information-intensive work \citep{eloundou2024gpts}, while also showing that exposure changes over time within broad occupation groups.

\begin{figure}[htbp]
    \vspace{0.3cm}
    \centering
    \includegraphics[width=\textwidth]{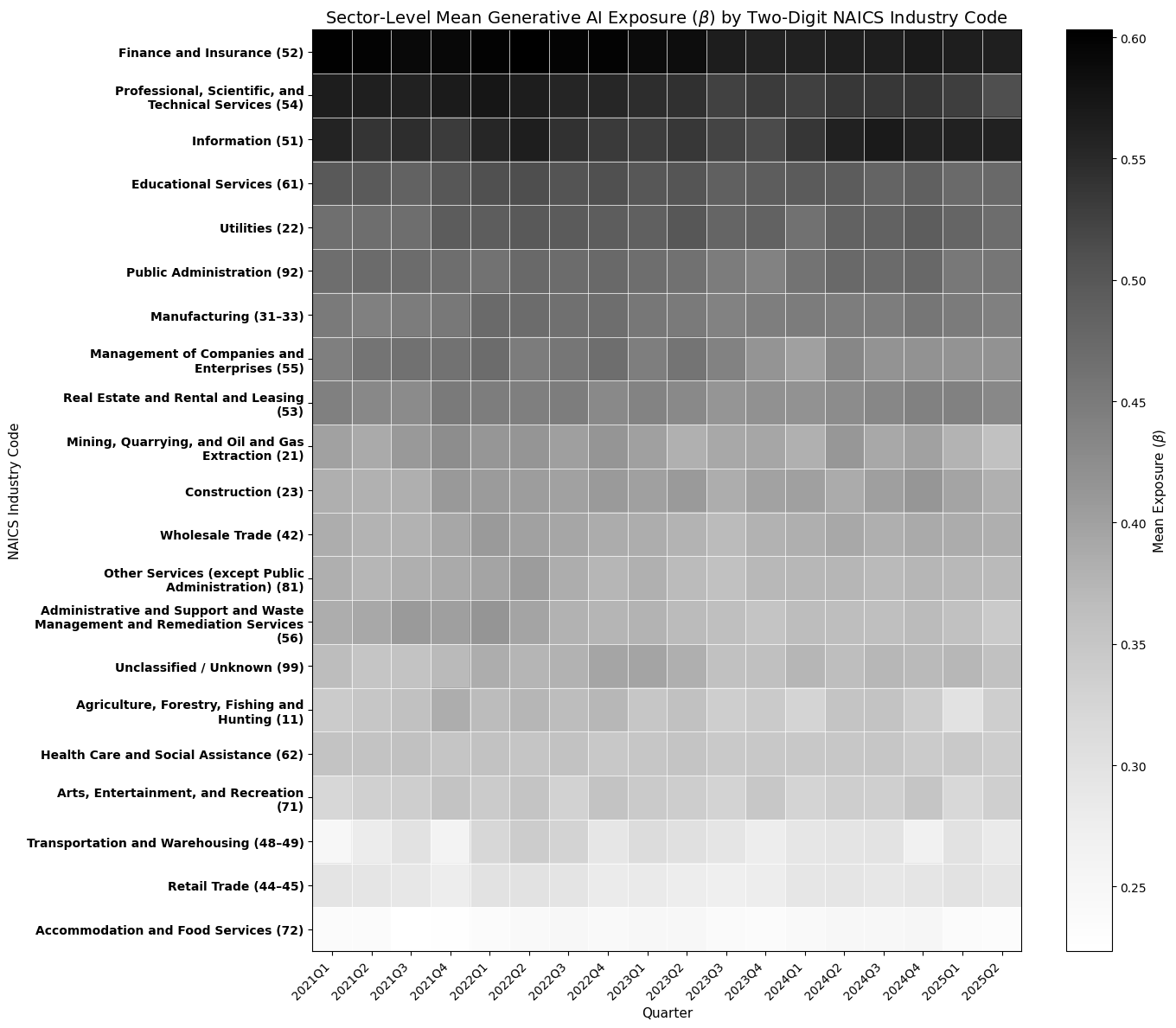}
    \caption{Sector-Level Mean Generative AI Exposure ($\beta$) by Two-Digit NAICS Industry Code}
    \label{fig:sector_heatmap_beta}
    \vspace{0.2cm}

    \begin{minipage}{\textwidth}
        \footnotesize
        \textit{Notes}: This figure reports mean posting-level exposure, measured by $\beta$, by two-digit NAICS sector and quarter. Darker colors indicate higher average exposure.
    \end{minipage}
\end{figure}

\textbf{Exposure across industries}: Figure~\ref{fig:sector_heatmap_beta} reports mean generative AI exposure by two-digit NAICS sector and quarter, with darker colors indicating higher average exposure. This figure uses the nationwide scope of the data to characterize how exposure is distributed across the full set of major industries. A detailed table of sector-level averages appears in Table~\ref{tab:sector_avg_exposure} in \ref{app:sector_avg_exposure}.


The heatmap reveals a clear cross-sector gradient. Exposure is highest in Finance and Insurance (NAICS 52, \(\bar{\beta}=0.584\)), Professional, Scientific, and Technical Services (NAICS 54, \(\bar{\beta}=0.548\)), and Information (NAICS 51, \(\bar{\beta}=0.545\)). It is lowest in Accommodation and Food Services (\(\bar{\beta}=0.239\)), Retail Trade (\(\bar{\beta}=0.289\)), and Transportation and Warehousing (\(\bar{\beta}=0.292\)). The gap between the top and bottom of the distribution is approximately 0.35 exposure units, indicating substantial heterogeneity in how generative AI maps onto posted job content across sectors.

This sectoral pattern is consistent with the nature of generative AI. Exposure is highest in sectors where posted work is more likely to involve language-intensive, information-processing, analytical, and digital tasks \citep{mckinsey2023state}. It is lowest in sectors where work more often requires physical presence, manual activity, or in-person service \citep{autor2003skill}. This does not imply that entire sectors are uniformly exposed or unexposed. Rather, the sectoral differences show how the task content of posted vacancies varies across the economy.

The heatmap also shows that exposure is not static within sectors. Many sectors exhibit quarter-to-quarter variation. High-exposure sectors tend to display more visible fluctuations, whereas lower-exposure sectors appear comparatively stable. These descriptive patterns reinforce the importance of a dynamic, posting-level measure: the incidence of generative AI exposure differs across sectors, but it also changes over time within sectors.

\section{Decomposition of Generative AI Exposure: Methods}
\label{sec:decomposition}

We next decompose changes in aggregate generative AI exposure to identify the margins through which labor demand changes over time. Our central distinction is between two forms of adjustment. Firms may change where they hire by reallocating postings across different types of jobs. They may also change what jobs contain by revising the task content of comparable jobs. The first margin is hiring reallocation; the second is job redesign.

Our main analysis uses a three-fold extension of the Kitagawa decomposition \citep{kitagawa1955components}. This decomposition expresses changes in aggregate exposure as the sum of three components: a composition effect, a within-cell exposure effect, and an interaction effect. The composition effect captures changes in the mix of posted jobs. The within-cell exposure effect captures changes in exposure within comparable jobs. The interaction effect captures the additional change that arises when these two adjustments occur simultaneously.

We complement the Kitagawa analysis with a regression-based Oaxaca--Blinder decomposition \citep{oaxaca1973male,blinder1973wage,oaxaca2025oaxaca}. The two decompositions are related but serve different purposes. The Kitagawa decomposition provides a cell-based accounting of aggregate exposure changes. The Oaxaca--Blinder decomposition instead asks which observable job characteristics, such as occupation, industry, seniority, location, remote-work status, internship status, and employment type, are most associated with the exposure gap between the pre- and post-GPT periods. Thus, Kitagawa identifies the adjustment margins, while Oaxaca--Blinder helps characterize the observable dimensions along which compositional change occurs.

\subsection{Cell-Level Representation of Aggregate Exposure}

Our exposure measure is defined at the posting level. To study aggregate labor-demand adjustment, we aggregate posting-level exposure to job cells. A cell is defined by occupation, seniority, and industry. This cell structure matches the main dimensions of our sampling strategy and captures three sources of heterogeneity central to our research design: the type of work being performed, the sector in which it is performed, and the career stage of the job.

Let \(\beta_p\) denote the posting-level exposure measure for posting \(p\), and let \(t\) index time periods. Aggregate exposure in period \(t\) is the average exposure across postings in that period:
\begin{equation*}
\bar{E}_t = \mathbb{E}[\beta_p \mid t].
\end{equation*}

Because every posting belongs to one mutually exclusive cell \(c\), aggregate exposure can be written as a weighted average of cell-level exposure:
\begin{equation}
\bar{E}_t
=
\sum_c
\mathbb{E}[\beta_p \mid c,t] \Pr(c \mid t).
\end{equation}

Define \(E_{c,t} \equiv \mathbb{E}[\beta_p \mid c,t]\) as mean exposure within cell \(c\) in period \(t\), and \(w_{c,t} \equiv \Pr(c \mid t)\) as the share of postings in period \(t\) that belong to cell \(c\). Then aggregate exposure can be written as
\begin{equation}
\bar{E}_t = \sum_c w_{c,t} E_{c,t}.
\label{eq:agg_exposure_cell}
\end{equation}

Equation~\eqref{eq:agg_exposure_cell} shows that aggregate exposure can change through two basic margins. First, firms may change the mix of jobs they post, which changes the weights \(w_{c,t}\). Second, the task content of comparable jobs may change over time, which changes \(E_{c,t}\). In our interpretation, the first margin captures hiring reallocation across job cells, while the second captures job redesign within cells.

Let period \(0\) denote the baseline period. In our main implementation, period \(0\) is the full year 2021. We use year 2021 as the baseline because it precedes the broad public diffusion of generative AI tools and provides a stable pre-GPT benchmark. Our goal is to decompose the change in aggregate exposure between period \(0\) and period \(t\):
\begin{equation}
\Delta \bar{E}_t = \bar{E}_t - \bar{E}_0.
\label{eq:agg_exposure_delta}
\end{equation}

\subsection{Three-Fold Kitagawa Decomposition}

We decompose the change in aggregate exposure into three counterfactual components:
\vspace{6pt}
\begin{equation}
\Delta \bar{E}_t
=
\underbrace{\sum_c (w_{c,t} - w_{c,0}) E_{c,0}}_{\text{Composition effect}}
+
\underbrace{\sum_c w_{c,0}(E_{c,t} - E_{c,0})}_{\text{Within-cell exposure effect}}
+
\underbrace{\sum_c (w_{c,t} - w_{c,0})(E_{c,t} - E_{c,0})}_{\text{Interaction effect}}.
\label{eq:threefold_main}
\vspace{9pt}
\end{equation}

The first term is the composition effect. It measures how aggregate exposure would change if posting shares shifted from their 2021 distribution to their period-\(t\) distribution, while exposure within each cell remained fixed at its 2021 level. Substantively, this term captures hiring reallocation across job cells. A positive value means that postings are shifting toward cells that were more exposed to generative AI in the baseline period. A negative value means that postings are shifting away from those more exposed cells.

The second term is the within-cell exposure effect. It measures how aggregate exposure would change if cell-level exposure shifted from its 2021 level to its period-\(t\) level, while posting shares remained fixed at their 2021 distribution. Substantively, this term captures job redesign within comparable jobs. A negative value indicates that the tasks described within a cell have become less exposed to generative AI over time, holding the baseline distribution of postings fixed. A positive value indicates movement toward more exposed task content within cells.

The third term is the interaction effect. It captures the additional change that arises because posting shares and within-cell exposure move at the same time. This term indicates whether hiring reallocation and job redesign reinforce or offset each other. For example, the interaction term is negative when postings shift toward cells whose exposure is falling, or away from cells whose exposure is rising. It is positive when the two movements push aggregate exposure in the same direction.

This three-fold formulation is useful in our setting because it separately identifies changes in where firms hire, changes in what comparable jobs contain, and the joint movement of the two. The classic two-fold Kitagawa decomposition allocates the interaction term into the composition and within-cell components. We report the corresponding two-fold decomposition in ~\ref{app:additional_decomp}.

\subsection{Common Support and Robustness of the Decomposition}

The set of observed cells can vary over time. Some occupation-by-seniority-by-industry cells may appear in the baseline period but not in a later period, or vice versa. To ensure that the decomposition compares comparable cells, we define the period-specific common support between the baseline year and period \(t\) as
\begin{equation*}
S_t = \{ c : w_{c,0} > 0 \text{ and } w_{c,t} > 0 \}.
\end{equation*}

Cells in \(S_t\) are observed in both 2021 and period \(t\). We re-normalize posting shares within this common-support sample before applying the decomposition. This approach focuses the analysis on changes among persistent job cells, rather than mechanically attributing changes to cell entry or exit. Because the re-normalized common-support aggregate is not identical to the raw aggregate, ~\ref{app:common_support} reports overlap diagnostics, reconstruction checks, and the gap between the raw and re-normalized aggregate series. These diagnostics show that the common-support series closely tracks the raw aggregate and that the reconstruction gap is negligible.

Our main text reports the three-fold decomposition in Equation~\eqref{eq:threefold_main}.~\ref{app:additional_decomp} reports two additional exercises. First, we implement the symmetric two-fold Kitagawa decomposition, which separates aggregate change into composition and within-cell components by allocating the interaction term evenly across the two. Second, we repeat the analysis using a balanced-cell sample restricted to cells observed throughout the full sample period. These analyses confirm that the main patterns are not driven by the treatment of the interaction term or by changes in cell support over time.

\subsection{Oaxaca--Blinder Decomposition}

The Kitagawa decomposition separates aggregate exposure changes into reallocation, redesign, and interaction components. It does not, however, identify which observable job characteristics are most associated with the exposure gap. For example, a large composition effect may reflect shifts across occupations, industries, seniority groups, locations, remote-work arrangements, or employment types. To examine these observable dimensions, we complement the Kitagawa analysis with a regression-based Oaxaca--Blinder decomposition \citep{oaxaca1973male,blinder1973wage,oaxaca2025oaxaca}.

The Oaxaca--Blinder decomposition compares two aggregate periods: the pre-GPT period and the post-GPT period. It decomposes the change in mean exposure into an explained component and an unexplained component. The explained component captures the portion associated with changes in observable job characteristics. The unexplained component captures the remaining portion associated with changes in the relationship between those characteristics and exposure, as well as unobserved factors.

We implement the decomposition at the cell level for computational tractability. Each cell is defined by a unique combination of occupation, seniority, and industry. The regression is weighted by the number of postings in each cell-period so that the estimates reproduce posting-level moments. The estimating equation is
\begin{equation}
Y_{c,g} = \alpha_g + X_{c,g}'\beta_g + \varepsilon_{c,g},
\end{equation}
where \(g \in \{A,B\}\) indexes the pre- and post-GPT periods, \(Y_{c,g}\) is mean exposure in cell \(c\) and period group \(g\), and \(X_{c,g}\) is a vector of observable job characteristics. The covariates are organized into blocks: occupation, industry, seniority, state, remote-work arrangement, internship status, and employment type. The occupation block includes 923 O*NET--SOC categories observed in our sample; the industry block includes 20 two-digit NAICS sectors; the seniority block includes junior, intermediate, and senior postings; the state block includes 51 state categories; the remote-work block includes remote, hybrid, and not-remote postings; the internship block distinguishes intern from non-intern postings; and the employment-type block includes full-time, part-time, and part-time/full-time postings. After omitting one reference category per block, the specification includes approximately 1{,}000 regressors.

Using the pre-GPT period as the reference structure, the two-fold decomposition is
\begin{equation}
\bar{Y}_B - \bar{Y}_A
=
\underbrace{(\bar{X}_B - \bar{X}_A)'\hat{\beta}_A}_{\text{Explained}}
+
\underbrace{(\hat{\alpha}_B - \hat{\alpha}_A)
+ \bar{X}_B'(\hat{\beta}_B - \hat{\beta}_A)}_{\text{Unexplained}}.
\vspace{9pt}
\end{equation}

The explained component answers a counterfactual question: how much would average exposure change if the distribution of observable job characteristics shifted from the pre-GPT period to the post-GPT period, while the association between those characteristics and exposure remained fixed at its pre-GPT level? It therefore captures the part of the exposure gap attributable to shifts in observed job characteristics.

This component is conceptually related to the Kitagawa composition effect because both involve changes in composition. However, the two objects are not identical in our implementation. The Kitagawa composition effect is a cell-based accounting term computed over the full set of occupation-by-industry-by-seniority cells. The Oaxaca--Blinder explained component is a regression-based decomposition that enters observable characteristics as additive covariate blocks. If the Oaxaca--Blinder specification were fully saturated with indicators for the exact same cells used in the Kitagawa decomposition, the explained component would closely correspond to the Kitagawa composition effect \citep{oaxaca2025oaxaca}. Our implementation instead uses the Oaxaca--Blinder decomposition as a complementary diagnostic tool: it identifies which observable characteristics account for the exposure change associated with compositional differences.

The unexplained component captures the remaining exposure gap after accounting for changes in observed job characteristics. Formally, it reflects changes in the intercept and in the estimated coefficients between the pre- and post-GPT periods. We do not interpret this component as a direct analog of the Kitagawa within-cell effect. Its magnitude depends on the choice of reference group, the included covariates, and the functional form of the regression \citep{fortin2011decomposition}. We therefore interpret it as the portion of the exposure gap not accounted for by shifts in observable job characteristics, rather than as a separate measure of job redesign.

We report detailed contributions only for the explained component and only at the level of covariate blocks. This choice follows standard practice in decomposition analysis. For categorical variables, the total contribution of a block, such as occupation or industry, is invariant to the choice of omitted reference category. In contrast, contributions of individual categories within a block depend on which category is omitted; changing the reference category can redistribute contributions across categories without changing the block total \citep{fortin2011decomposition}. We therefore report block-level explained contributions, such as the total contribution of occupation, industry, or seniority. For the unexplained component, category-level contributions are even more sensitive to the choice of reference group because they measure coefficient changes relative to the omitted category. We therefore interpret the unexplained component only in the aggregate.

\section{Results}
\label{sec:results}

This section presents the decomposition results. We first report the three-fold Kitagawa decomposition, which separates changes in aggregate generative AI exposure into hiring reallocation, within-cell job redesign, and their interaction. We then examine whether these margins differ across the job ladder. Finally, we report a complementary Oaxaca--Blinder decomposition to identify which observable job characteristics are most associated with the pre- versus post-GPT exposure gap.

\subsection{Aggregate Adjustment: Hiring Reallocation and Job Redesign}

\begin{figure}[htbp]
    \vspace{3pt}
    \centering
    \includegraphics[width=1\textwidth]{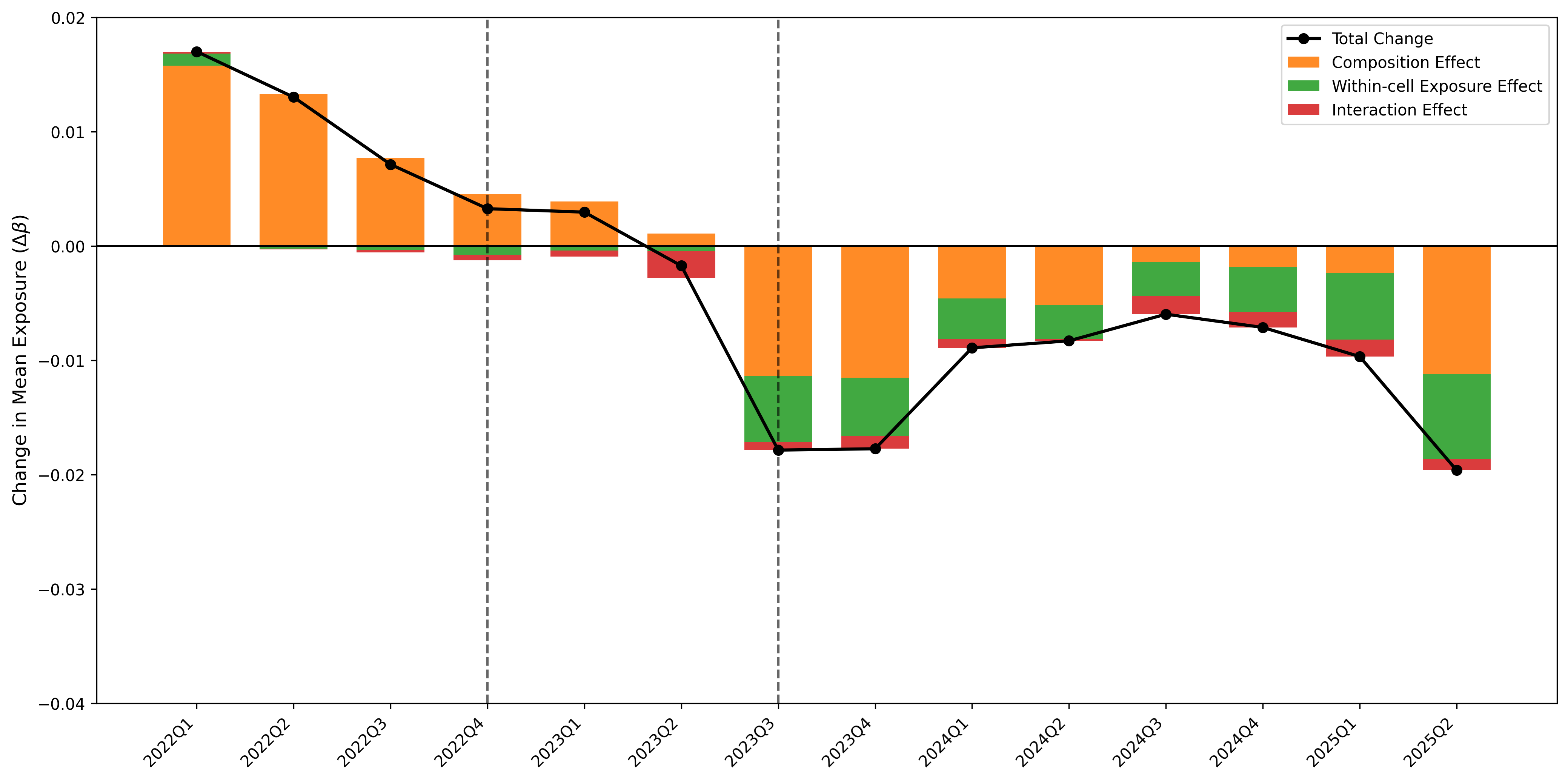}
    \caption{Three-Fold Decomposition of Changes in Aggregate Generative AI Exposure}
    \label{fig:overall_change}
    
    \begin{minipage}{1\textwidth}
    \vspace{0.5em}
    \footnotesize
    \textit{Notes:} The figure reports the three-fold Kitagawa decomposition of changes in aggregate exposure relative to the 2021 baseline. The black line shows the total change in aggregate exposure. Bars show the contribution of the composition effect, the within-cell exposure effect, and their interaction. 
    \end{minipage}
    \vspace{-6pt}
\end{figure}

Figure~\ref{fig:overall_change} presents the three-fold Kitagawa decomposition of changes in mean exposure relative to the 2021 baseline. The black line shows the total change in aggregate exposure. The bars separate this change into three components: a composition effect, a within-cell exposure effect, and an interaction effect. In substantive terms, the composition effect captures changes in where firms hire, the within-cell exposure effect captures changes in what comparable jobs contain, and the interaction effect captures the joint movement of these two margins.


Aggregate exposure rises relative to the 2021 baseline through early 2023, then turns negative beginning in Q3 of 2023 and declines through the end of the sample. The decomposition shows that this turning point reflects changes along both margins. Before Q3 of 2023, the composition effect is generally positive, indicating that the mix of postings shifts toward job cells that were more exposed at baseline. From that quarter onward, the composition effect becomes negative, indicating that posted hiring demand shifts away from cells that were more exposed in 2021. This pattern is consistent with a reallocation of hiring demand away from more exposed work.

The within-cell exposure effect follows a different trajectory. It remains close to zero before Q3 of 2023, then becomes increasingly negative. Because this component is measured within occupation-by-industry-by-seniority cells, it captures changes in exposure among comparable jobs rather than changes in the mix of broad job categories. Its decline therefore suggests that employers revise the task content of jobs they continue to post, reducing the exposure of those jobs to generative AI. This is the job-redesign margin that static occupation-level exposure measures cannot observe.

Table~\ref{tab:threefold_abs_contribution_post2023Q3} summarizes the relative importance of the three components from Q3 of 2023 onward. In the full sample, hiring reallocation accounts for 52.01\% of the aggregate absolute contribution. Within-cell job redesign accounts for 39.46\%, and the interaction effect accounts for 8.54\%. Thus, reallocation is the largest single margin, but redesign is also quantitatively important. The aggregate decline in exposure is therefore not simply a story of fewer postings in highly exposed occupations or sectors. A substantial share reflects changes in the task content of comparable jobs.


\begin{table}[htbp]
\centering
\caption{Relative Contributions in the Three-Fold Kitagawa Decomposition Since Q3 of 2023}
\vspace{-9pt}
\label{tab:threefold_abs_contribution_post2023Q3}

\begin{threeparttable}
\setlength{\tabcolsep}{12pt}
\begin{tabular}{lccc}
\toprule
 & Composition Effect & Within-cell Exposure Effect & Interaction Effect \\
\midrule
Overall      & 52.01 & 39.46 & 8.54 \\
Junior       & 60.15 & 18.22 & 21.63 \\
Intermediate & 47.66 & 45.02 & 7.32 \\
Senior       & 70.80 & 24.01 & 5.19 \\
\bottomrule
\end{tabular}

\begin{tablenotes}
\footnotesize
\item \textit{Notes}: This table reports the aggregate absolute contribution of each component in the three-fold Kitagawa decomposition of changes in generative AI exposure from Q3 of 2023 onward. Contributions are computed as
$\sum_t |X_t| \big/ \sum_t (|C_t| + |W_t| + |I_t|)$,
where $C_t$, $W_t$, and $I_t$ denote the composition, within-cell exposure, and interaction effects, respectively. Each row sums to 100\%.
\end{tablenotes}

\end{threeparttable}
\vspace{-3pt}
\end{table}

The timing of the within-cell decline is also informative. The shift begins around Q3 of 2023, shortly after enterprise-oriented generative AI tools became more widely available \citep{openai2023chatgptenterprise}. We do not interpret this timing as causal evidence of a specific product release. Rather, it is consistent with the broader idea that as generative AI became more deployable inside organizations, firms began to revise not only the allocation of vacancies but also the task requirements embedded in those vacancies.

The interaction effect is smaller than the other two components in the aggregate, but it is useful for understanding how reallocation and redesign move together. The interaction is generated when changes in posting shares and changes in within-cell exposure occur simultaneously within the same cell. A negative interaction can arise, for example, when firms expand hiring in cells whose exposure is falling, or when they reduce hiring in cells whose exposure is rising. A positive interaction arises when both movements push aggregate exposure in the same direction. Appendix~\ref{app:interaction_types_section} decomposes the interaction by sign configuration and shows that positive and negative configurations partially offset one another. This explains why the aggregate interaction term remains relatively small even though many cells experience simultaneous changes in posting shares and exposure.

Several additional analyses support the aggregate pattern. Appendices~\ref{app:overall_level} and~\ref{app:relative_contribution} report the corresponding exposure levels, counterfactual paths, and relative contributions over time. ~\ref{app:additional_decomp} reports the symmetric two-fold Kitagawa decomposition and the balanced-cell decomposition. These exercises yield qualitatively similar conclusions: the post-Q3-of-2023 decline in aggregate exposure reflects both reallocation across job cells and redesign within job cells.

\subsection{Heterogeneity Across the Job Ladder}

\begin{figure}[p]
    \centering
    \includegraphics[width=0.8\textwidth]{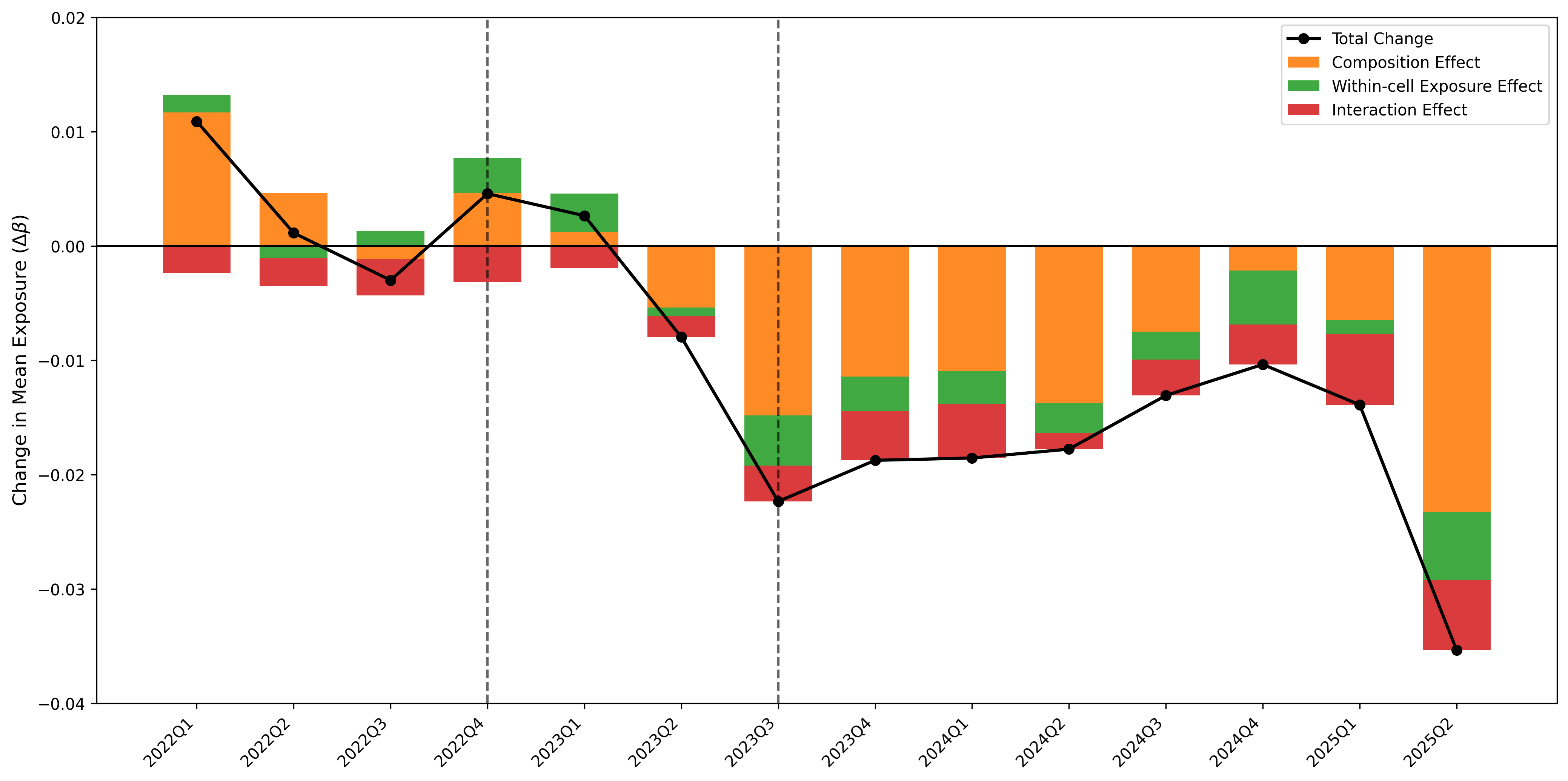}
    \vspace{-6pt}
    \caption{Three-Fold Decomposition of Changes in Generative AI Exposure: Junior Jobs}
    \label{fig:junior_change}
\end{figure}

\begin{figure}[p]
    \centering
    \includegraphics[width=0.8\textwidth]{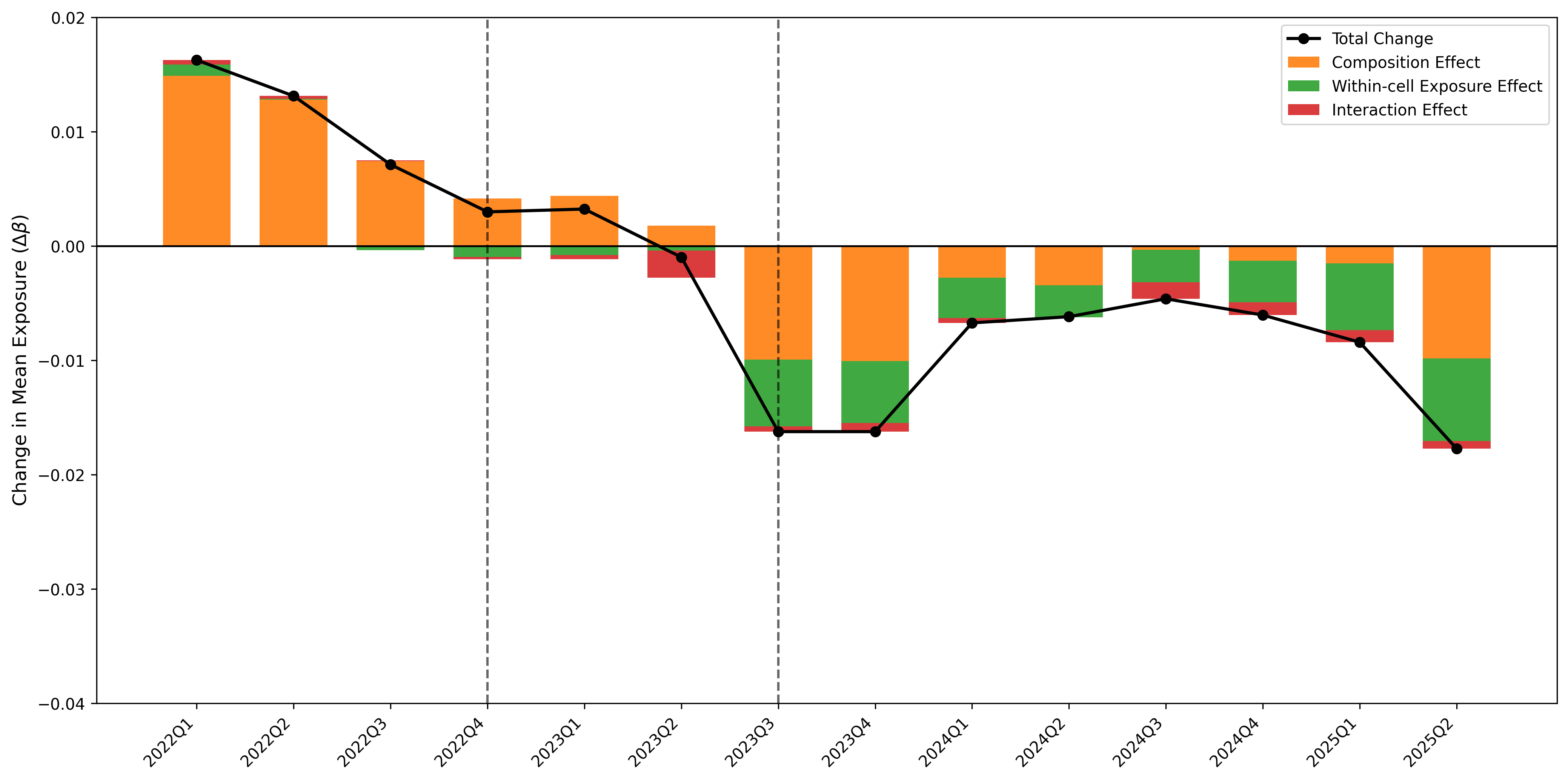}
    \vspace{-6pt}
    \caption{Three-Fold Decomposition of Changes in Generative AI Exposure: Intermediate Jobs}
    \label{fig:intermediate_change}
\end{figure}

\begin{figure}[p]
    \centering
    \includegraphics[width=0.8\textwidth]{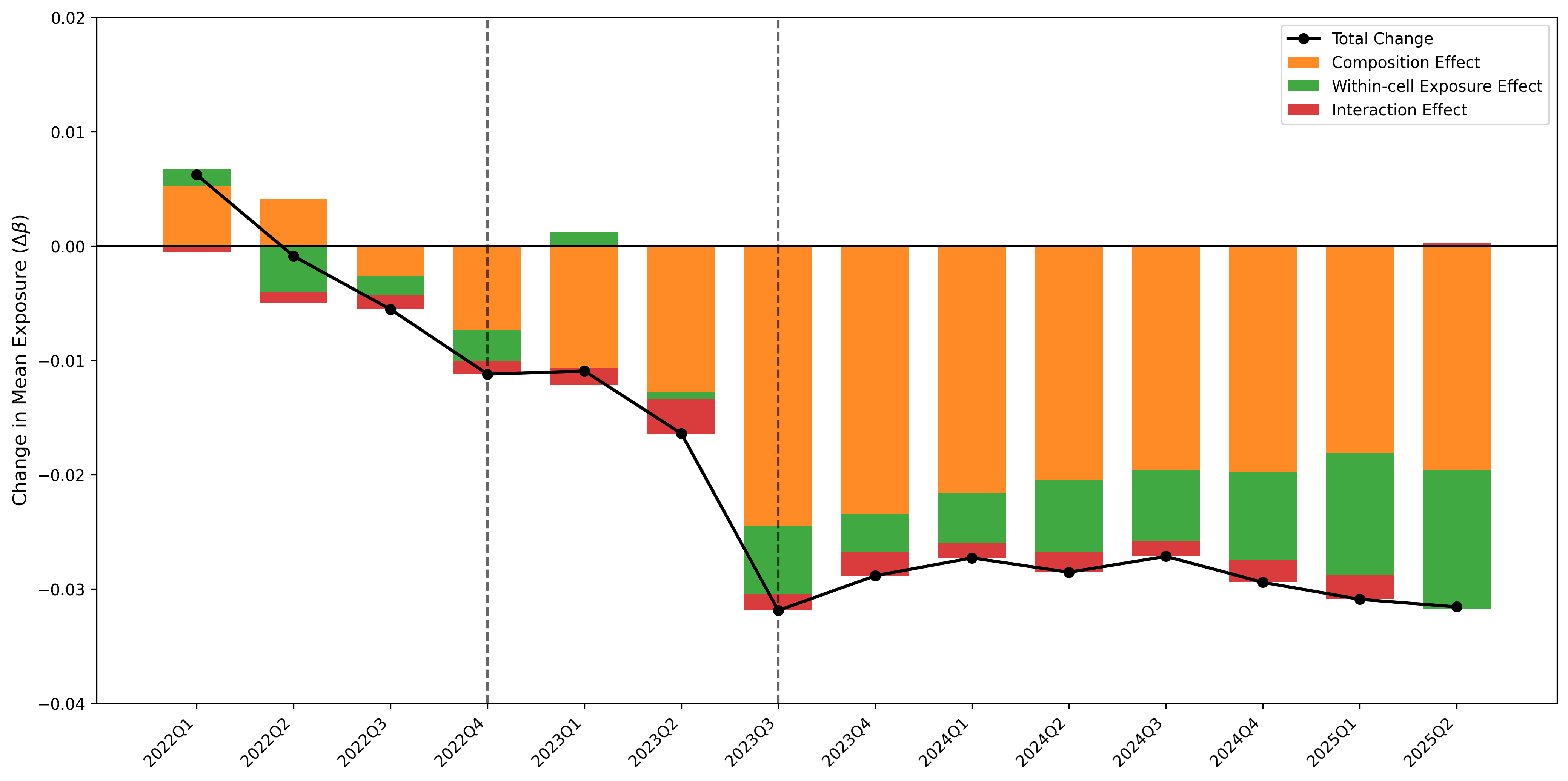}
    \vspace{-6pt}
    \caption{Three-Fold Decomposition of Changes in Generative AI Exposure: Senior Jobs}
    \label{fig:senior_change}
\end{figure}

We next examine whether the adjustment margins differ by job seniority. Figures~\ref{fig:junior_change}--\ref{fig:senior_change} repeat the three-fold Kitagawa decomposition separately for junior, intermediate, and senior postings. Table~\ref{tab:threefold_abs_contribution_post2023Q3} reports the aggregate absolute contribution of each component from Q3 of 2023 onward.


Junior postings adjust through all three components. From Q3 of 2023 onward, the composition effect accounts for 60.15\% of the aggregate absolute contribution, making hiring reallocation the largest margin. The within-cell exposure effect accounts for 18.22\%, while the interaction effect accounts for 21.63\%. The relatively large interaction term indicates that, for junior jobs, reallocation and redesign often occur together within cells. Appendix~\ref{app:junior_interaction} shows that the interaction is dominated by two negative configurations: cells in which junior hiring expands while exposure falls, and cells in which junior hiring contracts while exposure rises. This pattern suggests that entry-level work is not only reallocated across job categories but also reorganized within those categories.

Intermediate postings closely mirror the aggregate pattern. From Q3 of 2023 onward, the composition effect accounts for 47.66\% of the aggregate absolute contribution, while the within-cell exposure effect accounts for 45.02\%. The interaction effect accounts for only 7.32\%. This indicates that intermediate jobs adjust through both hiring reallocation and job redesign, with the two margins contributing almost equally.

Senior postings exhibit a distinct sequence of adjustment. The composition effect turns negative earlier than in the aggregate series, beginning in Q2 of 2022, and remains large through the end of the sample. From Q3 of 2023 onward, the composition effect accounts for 70.80\% of the aggregate absolute contribution. The within-cell exposure effect becomes more clearly negative after Q3 of 2023 and accounts for 24.01\%, while the interaction effect accounts for only 5.19\%. Thus, senior jobs adjust primarily through reallocation across senior vacancies, with within-cell redesign emerging later and playing a secondary role.

These seniority patterns clarify our contribution to the literature on generative AI and the job ladder. Much of the existing debate asks whether junior workers are more exposed or more adversely affected than senior workers. Our results show that the relevant heterogeneity is not only about the magnitude of adjustment, but also about the margin of adjustment. Senior jobs adjust earlier and mainly through reallocation. Junior jobs adjust through a broader mix of reallocation, redesign, and their interaction. Generative AI therefore does not simply move labor demand up or down the job ladder. It changes how different layers of the hierarchy absorb technological change.

\subsection{Robustness to Cross-Sector Reallocation}

A concern in interpreting changes in exposed labor demand is that AI exposure may be correlated with macroeconomic sensitivity. For example, highly exposed work is often located in sectors such as information, finance, and professional services, which may also be more sensitive to monetary tightening and other aggregate shocks \citep{iscenko2026looking}. If aggregate exposure declines because hiring shifts away from these sectors, the decline could partly reflect macroeconomic conditions rather than adjustment to generative AI.

To address this concern, we implement a within-sector version of the Kitagawa decomposition. We run the decomposition separately within each two-digit NAICS sector and then aggregate the sector-specific components using fixed baseline sector weights. This procedure removes cross-sector reallocation as a source of aggregate exposure change and therefore asks whether the main results persist within sectors.

The results, reported in ~\ref{app:within_sector}, are consistent with the baseline decomposition. The within-cell exposure effect remains similar in timing and magnitude, suggesting that the job-redesign margin is not driven by cross-sector shifts. The composition effect is smaller once cross-sector variation is removed, as expected, but it remains economically meaningful and turns negative from Q3 of 2023 onward. These results indicate that sector-level hiring dynamics do not fully account for the observed adjustment. Posted labor demand changes through both reallocation and redesign even when comparisons are restricted within sectors.

\subsection{Oaxaca--Blinder Decomposition: Observable Sources of the Exposure Gap}

The Kitagawa results identify the margins through which aggregate exposure changes. We now use the Oaxaca--Blinder decomposition to examine which observable job characteristics are associated with the pre- versus post-GPT exposure gap. This analysis is complementary to the Kitagawa decomposition. It should not be interpreted as mechanically decomposing the Kitagawa composition effect. Instead, it provides a regression-based accounting of how much of the exposure gap is associated with shifts in observed characteristics such as occupation, industry, seniority, location, remote-work arrangement, internship status, and employment type.

Table~\ref{tab:ob_combined} reports the weighted Oaxaca--Blinder decomposition for the full sample and separately by seniority. Among all jobs, average exposure falls from 0.404 in the pre-GPT period to 0.389 in the post-GPT period, a decline of 0.015. The explained component is $-0.010$, accounting for about two-thirds of the decline. The unexplained component is $-0.005$, accounting for the remaining third. Thus, both observed compositional shifts and changes in exposure conditional on observed characteristics contribute to the post-GPT decline.


\begin{table}[htbp]
\centering
\caption{Weighted Oaxaca--Blinder Decomposition of Aggregate Exposure: Pre- vs.\ Post-GPT}
\vspace{-9pt}
\label{tab:ob_combined}
\begin{threeparttable}
\setlength{\tabcolsep}{12.5pt}
\begin{tabular}{lcccc}
\toprule
 & Pre-GPT mean & Post-GPT mean & Explained & Unexplained \\
\midrule
Overall            & 0.404 & 0.389 & $-$0.010 & $-$0.005 \\
Junior jobs        & 0.435 & 0.419 & $-$0.011 & $-$0.006 \\
Intermediate jobs  & 0.388 & 0.374 & $-$0.008 & $-$0.005 \\
Senior jobs        & 0.595 & 0.571 & $-$0.018 & $-$0.006 \\
\bottomrule
\end{tabular}
\begin{tablenotes}
\scriptsize
\item \textit{Notes:} The first two columns report weighted average exposure in the pre-GPT and post-GPT periods. The last two columns report the weighted Oaxaca--Blinder decomposition of the pre-post difference, estimated using separate weighted regressions for the two periods. 
\end{tablenotes}
\end{threeparttable}
\vspace{-6pt}
\end{table}

The same qualitative pattern appears across the job ladder. Average exposure declines for junior, intermediate, and senior jobs, and both the explained and unexplained components are negative in each group. The largest decline occurs among senior jobs, where exposure falls from 0.595 to 0.571. Senior jobs also have the largest explained component in absolute value, $-0.018$, which accounts for approximately 75\% of the total decline. For junior and intermediate jobs, the explained component accounts for a smaller but still substantial share of the decline. These results reinforce the Kitagawa evidence that compositional adjustment is especially pronounced among senior postings, while also showing that exposure declines conditional on observed characteristics across all seniority groups.

\subsection{Which Job Characteristics Account for the Explained Component?}

\begin{figure}[htbp]
    \centering
    \includegraphics[width=0.8\textwidth]{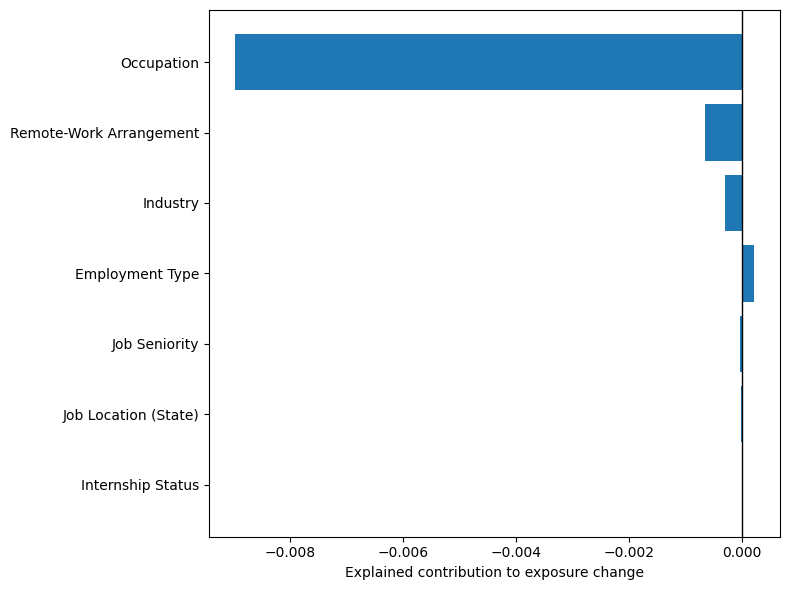}
    \vspace{-6pt}
    \caption{Explained Component by Observed Job-Characteristic Block: Pre-GPT vs.\ Post-GPT}
    \label{fig:ob_explained_blocks}
\end{figure}

Figure~\ref{fig:ob_explained_blocks} decomposes the explained component into contributions from observed job-characteristic blocks. Negative values indicate that shifts in the distribution of a characteristic are associated with lower post-GPT exposure, holding the pre-GPT association between that characteristic and exposure fixed.


The dominant observable source of the explained decline is occupation. Shifts in occupational composition account for about 90\% of the exposure change attributable to observed job characteristics. This means that, among the characteristics included in the Oaxaca--Blinder model, changes in the distribution of postings across occupations are the largest contributor to the post-GPT exposure decline. This result is consistent with prior occupation-level analyses showing that generative AI exposure varies substantially across occupations.

Other blocks contribute more modestly and in different directions. Remote-work arrangement contributes negatively: the share of remote postings declines from 6.1\% before GPT to 5.1\% after GPT, and remote postings have much higher pre-GPT exposure than non-remote postings (0.645 vs.\ 0.386). This shift away from remote work therefore lowers aggregate exposure, consistent with recent evidence linking generative AI exposure to reduced remote hiring, although broader return-to-office and labor-market dynamics may also contribute \citep{schubert2025organizational}. Industry also contributes negatively, reflecting some movement toward lower-exposure sectors. Employment type moves in the opposite direction. The share of full-time postings rises slightly from 82.3\% to 82.5\%, and full-time postings have higher pre-GPT exposure than part-time postings (0.434 vs.\ 0.271). This shift toward full-time jobs therefore increases predicted exposure and partially offsets the overall decline. Contributions from seniority, internship status, and location are negative but small.

These results sharpen the interpretation of the compositional adjustment. The dominant observable dimension is occupation, but the adjustment is not occupation-only. Posting-level characteristics such as remote-work arrangement, sector, and employment type also shape the exposure gap. This is one reason the posting-level data are useful: they allow us to observe dimensions of vacancy design that are not available in standard occupation-level exposure measures.

\subsection{Observable Sources of the Exposure Gap by Seniority}

\begin{figure}[htbp]
    \centering
    \vspace{6pt}
    \includegraphics[width=0.8\textwidth]{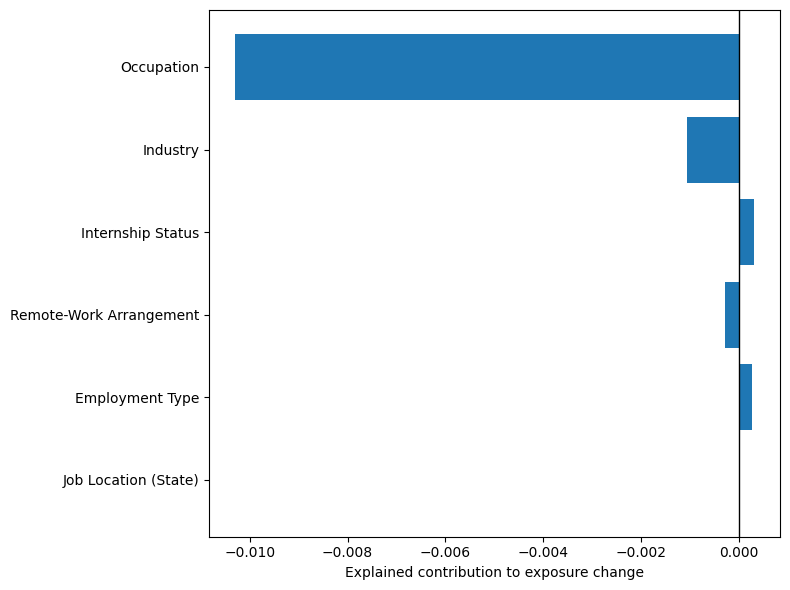}
    \vspace{-6pt}
    \caption{Explained Component by Observed Job-Characteristic Block for Junior Jobs}
    \label{fig:ob_explained_blocks_junior}
\end{figure}

\begin{figure}[htbp]
    \centering
    \includegraphics[width=0.8\textwidth]{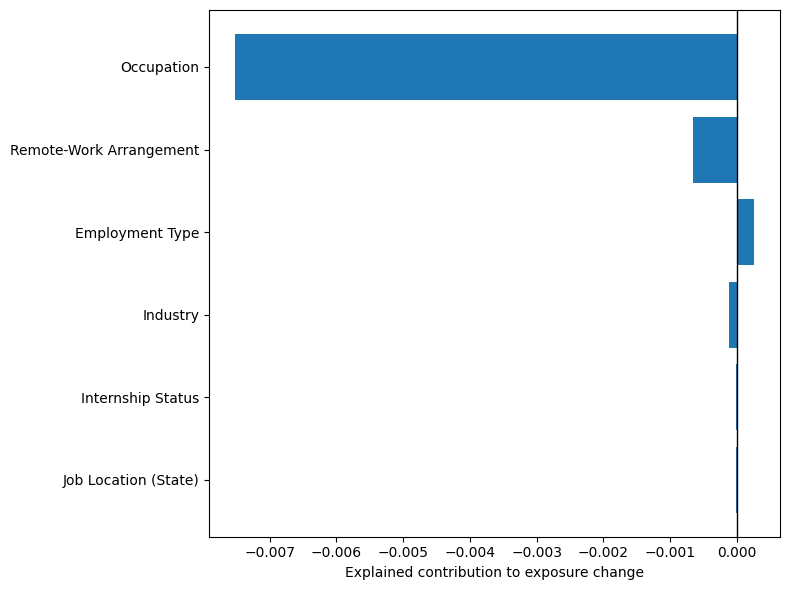}
    \vspace{-6pt}
    \caption{Explained Component by Observed Job-Characteristic Block for Intermediate Jobs}
    \label{fig:ob_explained_blocks_intermediate}
\end{figure}

\begin{figure}[htbp]
    \centering
    \includegraphics[width=0.8\textwidth]{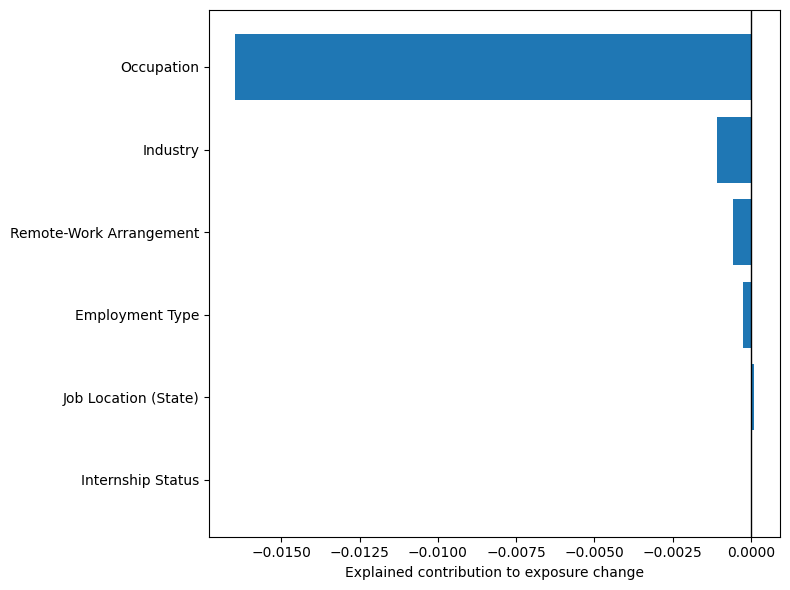}
    \vspace{-6pt}
    \caption{Explained Component by Observed Job-Characteristic Block for Senior Jobs}
    \label{fig:ob_explained_blocks_senior}
\end{figure}

Figures~\ref{fig:ob_explained_blocks_junior}--\ref{fig:ob_explained_blocks_senior} repeat the block-level Oaxaca--Blinder decomposition separately for junior, intermediate, and senior postings. Occupation is the dominant negative contributor at every seniority level, indicating that occupational reallocation is a common feature of post-GPT adjustment across the job ladder. The secondary margins, however, differ across seniority groups.


For junior jobs, industry is the second-largest negative contributor, indicating that sectoral reallocation toward lower-exposure industries is more visible at the entry level than in the aggregate. Internship status and employment type move in the opposite direction. The internship share rises from 10.2\% to 11.9\%, and internship postings have higher pre-GPT exposure than non-internship postings (0.581 vs.\ 0.419). Similarly, the full-time share rises slightly from 86.8\% to 87.1\%, and full-time junior postings have higher pre-GPT exposure than part-time or mixed postings (0.451 vs.\ 0.359 and 0.306). These shifts partially offset the exposure-reducing effects of occupational and sectoral reallocation. Remote-work arrangement contributes slightly negatively: the decline in remote postings from 5.4\% to 5.2\% more than offsets the rise in hybrid postings from 1.2\% to 2.0\%, and both remote and hybrid postings are more exposed than non-remote postings under the pre-GPT structure. Overall, junior jobs show a mixed compositional response, with occupational and sectoral shifts lowering exposure while internship and employment type push exposure upward.

Intermediate jobs closely mirror the aggregate pattern. Occupation accounts for approximately 90\% of the explained component, while industry and remote-work arrangement contribute modestly in the negative direction and employment type contributes slightly in the positive direction.

For senior jobs, occupation again dominates, but industry and remote-work arrangement are more visibly negative than in the other groups. The remote posting share falls from 15.2\% before GPT to 13.0\% after GPT, a larger decline than for junior or intermediate jobs. Because remote and hybrid senior postings have substantially higher pre-GPT exposure than non-remote senior postings (0.665 and 0.652 vs.\ 0.580), this shift away from remote senior vacancies lowers aggregate exposure within the senior group. This pattern is consistent with evidence that firms adopting generative AI reduce their remote-work share after ChatGPT \citep{schubert2025organizational}, and our results show that this margin is especially relevant among senior postings. Employment type also contributes negatively for senior jobs: the full-time share declines from 96.1\% to 95.0\%, and full-time senior postings have higher pre-GPT exposure than part-time or mixed senior postings (0.600 vs.\ 0.438 and 0.502). Thus, unlike junior and intermediate jobs, senior jobs exhibit a more uniformly exposure-reducing compositional response, with occupational shifts, remote-work changes, and employment-type changes all moving in the same direction.

Taken together, the Oaxaca--Blinder results show that occupational reallocation is the common core of the explained exposure decline, while secondary sources of compositional change differ across the job ladder. This complements the Kitagawa results. The Kitagawa decomposition shows that junior, intermediate, and senior jobs adjust through different combinations of reallocation and redesign. The Oaxaca--Blinder decomposition shows that the observable dimensions of compositional adjustment also differ by seniority. Generative AI-related labor-demand adjustment therefore involves both a broad occupational shift and a seniority-specific reorganization of the kinds of vacancies firms post.

\section{Conclusions}
\label{sec:conclusion}

This paper studies how generative AI is associated with the reorganization of labor demand. Using a nationwide sample of more than nine million job postings in the United States from 2021 through 2025, we construct a dynamic, posting-level measure of generative AI exposure with a two-stage large language model pipeline. We then decompose changes in aggregate exposure into two margins of adjustment: hiring reallocation across job cells and task redesign within comparable jobs. This approach allows us to move beyond the question of whether more exposed occupations decline and examine how the structure and content of labor demand change as generative AI diffuses.

The results show that exposure to generative AI is not a fixed attribute of occupations. It varies across postings within the same occupation, differs systematically by industry and seniority, and changes over time. This finding is important because much of the emerging evidence on AI and labor markets relies on static, occupation-level exposure measures. Such measures are valuable for identifying where generative AI may matter ex ante, but they cannot observe whether employers subsequently revise the task content of jobs. Our evidence suggests that this revision margin is empirically meaningful.

The decomposition results show that the decline in aggregate exposure after the third quarter of 2023 reflects both changes in where firms hire and changes in what comparable jobs contain. Hiring reallocation is the largest single margin, accounting for 52.01\% of the aggregate absolute contribution from Q3 of 2023 onward. Within-cell task redesign accounts for 39.46\%, while the interaction between the two margins accounts for the remaining 8.54\%. Thus, the post-diffusion decline in exposure is not simply a story of fewer postings in highly exposed occupations, industries, or seniority groups. A substantial portion reflects changes in the task content of jobs that firms continue to post. The timing of this shift is consistent with the broader organizational diffusion of generative AI tools, although our design does not attribute the break to any single product release or adoption event.

A complementary Oaxaca--Blinder decomposition further clarifies the observable dimensions of compositional adjustment. Shifts in occupational composition account for most of the exposure change attributable to observed job characteristics. This result is consistent with occupation-level studies showing that generative AI exposure differs sharply across types of work. At the same time, other posting-level characteristics, including industry, remote-work arrangement, employment type, and internship status, also contribute to the exposure gap. These results reinforce the value of job-posting data: they reveal labor-demand adjustments along dimensions that are difficult to observe in occupation-level measures alone.

The adjustment patterns also differ across the job ladder. Senior jobs adjust earlier and primarily through hiring reallocation. Junior jobs exhibit a broader pattern, with hiring reallocation, within-cell task redesign, and their interaction all contributing meaningfully. These results suggest that generative AI does not simply affect some career stages more than others. It also changes the mechanisms through which different levels of the job hierarchy absorb technological change. For junior jobs, the simultaneous movement of hiring reallocation and task redesign is especially important because entry-level positions are a major channel through which workers acquire skills and progress into more advanced roles.

These findings have three broader implications. First, they suggest that the labor-market effects of generative AI should be understood as a process of organizational reconfiguration, not only as a process of job displacement or augmentation. Firms appear to adjust both the allocation of hiring demand and the task architecture of posted jobs. Second, they show that exposure to a general-purpose technology can be endogenous to organizational adaptation. As firms learn about and implement generative AI, the task content of jobs may change, which means that exposure measured at one point in time may not fully describe exposure later in the diffusion process. Third, the results imply that the job ladder is an important site of adjustment. Changes in junior jobs may affect not only current hiring patterns but also the structure of early-career learning opportunities.

The paper also has implications for future research. Studies that link AI exposure to employment, wages, or worker mobility should account for the possibility that exposure itself changes as firms adapt. Static occupation-level exposure measures may miss within-occupation task redesign and may understate heterogeneity across seniority levels, sectors, and work arrangements. Dynamic, text-based measures can complement existing approaches by capturing how employers describe work in real time. More broadly, the results suggest that understanding the future of work requires measuring not only which jobs are exposed to AI, but also how jobs are being rewritten as AI capabilities diffuse.

Several limitations point to directions for future work. Job postings capture employers' stated demand at the point of hiring, but they do not directly measure realized work inside firms or the experiences of incumbent workers. Linking posting-based exposure measures to matched employer--employee data would make it possible to study how hiring reallocation and task redesign translate into employment, wage, and career outcomes. Future work could also examine firm-level heterogeneity in adjustment, including whether early adopters, large firms, or firms in AI-intensive sectors redesign jobs differently. Finally, as generative AI capabilities continue to advance and agentic AI systems become more widely deployed, tracking exposure and adjustment in real time will remain important for understanding the long-run consequences of this technology for labor markets and organizations.

\clearpage

\doublespacing

\bibliographystyle{aea}
\bibliography{references}

\clearpage

\pagenumbering{arabic}
\setcounter{page}{1}

\appendix
\renewcommand{\thesection}{Appendix~\Alph{section}}
\renewcommand{\thesubsection}{\Alph{section}.\arabic{subsection}}
\setcounter{table}{0}
\setcounter{figure}{0}
\renewcommand{\thetable}{\Alph{section}\arabic{table}}
\renewcommand{\thefigure}{\Alph{section}\arabic{figure}}

\begin{center}
    {\Large \textbf{Appendices for \\ ``Generative AI and the Reorganization of Labor Demand''}}
    
    \vspace{0.5cm}
    
    {\large Fangyan Wang, Zaiyan Wei, Yang Wang}
    
    
    {\small Mitch Daniels School of Business, Purdue University; West Lafayette, IN 47907}
\end{center}



\section{Full Annotation Prompts}
\label{app:full_prompts}

This appendix reproduces the full prompts used in the two-stage annotation pipeline. Stage 1 uses \texttt{Llama-3.1-8b-instant} for task extraction and task--skill matching. Stage 2 uses \texttt{GPT-5-nano} for task-level exposure classification.

\subsection{Stage 1 Prompt: Task Extraction and Task--Skill Matching}
\label{app:prompt_stage1}

\begin{PromptBlock}

You are an expert in job task analysis and workplace activities.

Your goal is to extract REAL, POSTING-SPECIFIC tasks from individual job postings.
These tasks will later be used to evaluate generative-AI exposure, so accuracy and realism are critical.

Analyze each job posting and output EXACTLY ONE valid JSON object matching the schema below.
Return ONLY JSON — no markdown, no notes, no explanations.

Each posting provides:
- ID: unique identifier
- TITLE_NAME: job title
- BODY: job description text
- SPECIALIZED_SKILLS_NAME: list of technical / hard skills
- COMMON_SKILLS_NAME: list of general / soft skills

IMPORTANT GUIDANCE:
- You are NOT summarizing an occupation or a typical role.
- You are extracting what THIS specific job posting describes workers actually doing.
- Do NOT generalize, normalize, or rewrite tasks into abstract or generic duties.
- Do NOT invent tasks that are not grounded in the posting text.
- Tasks should be traceable to the BODY text (verbatim or lightly edited for clarity).

STEP 1. Group Skills (Supportive Context Only)
- Group skills by semantic similarity within their original type.
- Specialized skills → S1, S2, ...
- Common skills → C1, C2, ...
- Each group: {"group_id": "...", "group_skills": [up to 5 similar skills]}.
- Do NOT mix specialized and common skills in the same group.
- If no skills are provided, create exactly one group:
  {"group_id": "NS0", "group_skills": []}.

STEP 2. Extract Posting-Specific Tasks (HIGH FIDELITY)
Extract 3–10 concrete, actionable tasks describing what the worker ACTUALLY DOES in this posting.

Task extraction rules:
- Tasks must be grounded in the BODY text and reflect real actions, tools, systems,
  materials, interactions, or work settings mentioned.
- Preserve specificity whenever possible (software names, equipment, documents,
  samples, customers, patients, systems, environments).
- Focus on observable actions (clear verbs + objects).
- Only include tasks explicitly stated or clearly implied by the BODY text.

STRICTLY AVOID:
- Company information, compensation, benefits, qualifications, or hiring instructions.

--------------------------------------------------
STEP 3. Match Tasks to Skill Groups
--------------------------------------------------
- Assign EACH task exactly ONE skill_group_id.
- Compare the task against all skill groups (specialized + common).
- Choose the closest semantic match.
- In case of ties, prefer specialized groups (S*) over common groups (C*).
- If no skills exist, assign all tasks to NS0.

--------------------------------------------------
OUTPUT FORMAT (STRICT — JSON ONLY)
--------------------------------------------------
Return EXACTLY one JSON object with this schema:

{
  "posting_id": "<ID>",
  "posting_title": "<TITLE_NAME>",
  "skills_groups": [
    {"group_id": "S1", "group_skills": ["..."]},
    {"group_id": "C1", "group_skills": ["..."]},
    {"group_id": "NS0", "group_skills": []}
  ],
  "tasks": [
    {
      "task_id": "t1",
      "task": "<8–50-word posting-specific task>",
      "skill_group_id": "S1 | C1 | NS0"
    }
  ]
}

FINAL CHECKS:
- Valid JSON only.
- 3–10 tasks.
- Tasks reflect concrete work described in the posting, not generalized role descriptions.
- Every task has exactly one skill_group_id.
\end{PromptBlock}

\subsection{Stage 2 Prompt: Exposure Classification}
\label{app:prompt_stage2}

\begin{PromptBlock}

You are an expert on Generative AI. Your task is to classify work tasks by their exposure to GenAI/LLM tools.
Assume:
- A worker with average expertise in this role.
- Access to powerful LLMs and readily available GenAI tools.
- Access to standard laptop tools (e.g., microphone, speakers).
- NO physical tools or physical presence beyond a laptop.

You will receive ONE JSON object containing:
- posting_id: string. Job posting ID.
- posting_title: string. Job title.
- tasks: list of task objects, each with:
    - task_id: string (e.g., "t1")
    - task: string describing the worker task activity.

Assign exactly ONE exposure_label to each task in { "E0", "E1", "E2" }:

E0 (No exposure):
- Label E0 if readily available GenAI/LLM tools (ChatGPT, copilots, ASR/TTS, multimodal tools) cannot reduce task time by ≥50% at equivalent quality.
- Tasks requiring substantial in-person interaction, physical manipulation, inspections, repairs, hands-on care, or credential-bound decisions must be E0.

E1 (Direct exposure):
- Label E1 if a single off-the-shelf GenAI/LLM tool, with no special integrations or fine-tuning, can reduce effort by ≥50% at equivalent quality.
-   Typical E1 patterns (not exhaustive):
  * Writing or transforming text or code
  * Translation, tone/style edits
  * Summarizing medium-length documents; extracting structured info
  * Drafting emails, briefs, slides, Q&A based on provided content
  * ASR transcription; TTS draft narration; simple audio clean-up
  * Simple image generation/editing (thumbnails, captions, marketing images)
- Not E1: tasks needing internal data access, enterprise systems, custom workflows, or automation beyond copy-paste.

E2 (LLM+ exposure):
- Label E2 if a single off-the-shelf GenAI/LLM tool alone cannot achieve ≥50% time reduction, but a thin AI-powered software layer could plausibly do so.
- Examples of thin layers:
  * Retrieval or search over private/internal documents
  * Multi-step workflows (draft → review → format → upload)
  * In-product assist (CRM, IDE, helpdesk) that reads context and suggests actions/replies.
  * Auto-suggest replies in customer service systems
  * Rubric/policy/brand enforcement; light fine-tuning
  * Integrated multimodal tools tied to asset libraries or translation pipelines
- Not E2:
  * If E1 already applies → choose E1
  * If even with integration AI cannot plausibly halve time → choose E0
  * If heavy bespoke ML, robotics, or new model training is required → choose E0

OUTPUT FORMAT (STRICT):
Return ONE JSON object:

{
  "posting_id": "<same as input>",
  "task_exposures": [
    {"task_id": "t1", "exposure_label": "E1"},
    {"task_id": "t2", "exposure_label": "E0"}
  ]
}

Rules:
- Return ONLY valid JSON. No explanation or extra text.
- Every input task_id must appear exactly once.
- exposure_label must be exactly one of "E0", "E1", or "E2".


\end{PromptBlock}

\section{Comparison with Eloundou et al.~(2024) }
\label{app:compare_eloundou}

To benchmark our measure against the existing literature, we compare our posting-based exposure measure with the occupation-level exposure measure in \citet{eloundou2024gpts}. Their measure is constructed from O*NET task data and assigns each occupation a single, time-invariant exposure score, whereas our measure is constructed from job postings and can vary across postings within the same occupation and over time.

\begin{figure}[htbp]
    \centering
    \includegraphics[width=0.7\textwidth]{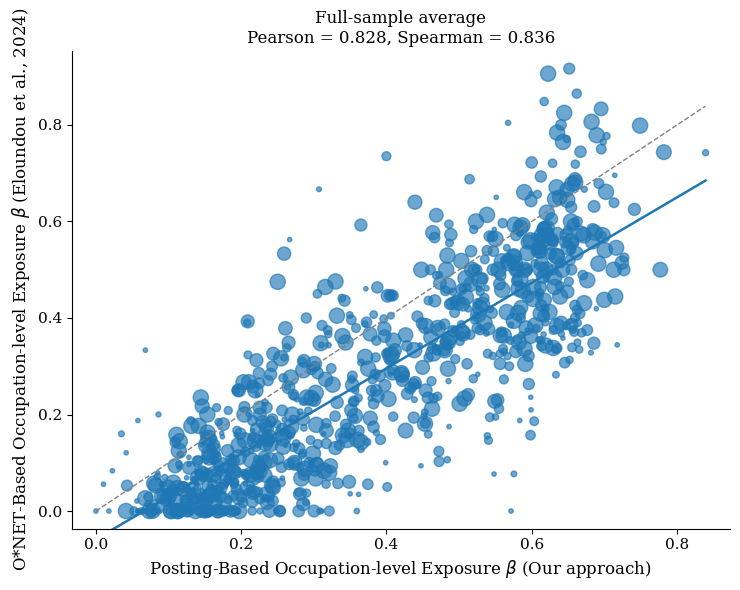}
    \caption{Comparison with the Occupation-Level Exposure Measure in Eloundou et al.~(2024)}
    \label{fig:eloundou_compare_appendix}

    \begin{minipage}{1\textwidth}
        \footnotesize
        \textit{Notes}: This figure compares our posting-based occupation-level exposure measure with the occupation-level exposure measure in \citet{eloundou2024gpts}. Our measure is aggregated from the posting level to the occupation level using the full sample. Each point represents a matched occupation.
    \end{minipage}
\end{figure}

To make the two approaches comparable, we aggregate our posting-level exposure measure \(\beta_p\) to the occupation level and compare it with the benchmark in \citet{eloundou2024gpts}. Figure~\ref{fig:eloundou_compare_appendix} shows that the full-sample occupation-level average of our measure is highly correlated with the benchmark in \citet{eloundou2024gpts}. Thus, at the aggregate occupation level, our measure is broadly aligned with the benchmark in \citet{eloundou2024gpts}.

However, this aggregate comparison masks two dimensions of heterogeneity that a single occupation-level score cannot capture. First, exposure differs across seniority levels within the same occupation. Second, occupation-level exposure itself evolves over time. These additional margins are consistent with the broader descriptive evidence in Appendix Figures~\ref{fig:e0e1e2_appendix}--\ref{fig:occ_terciles_seniority_appendix}, which show that exposure varies across occupation groups, seniority levels, and time.

\begin{figure}[htbp]
    \centering
    \includegraphics[width=\textwidth]{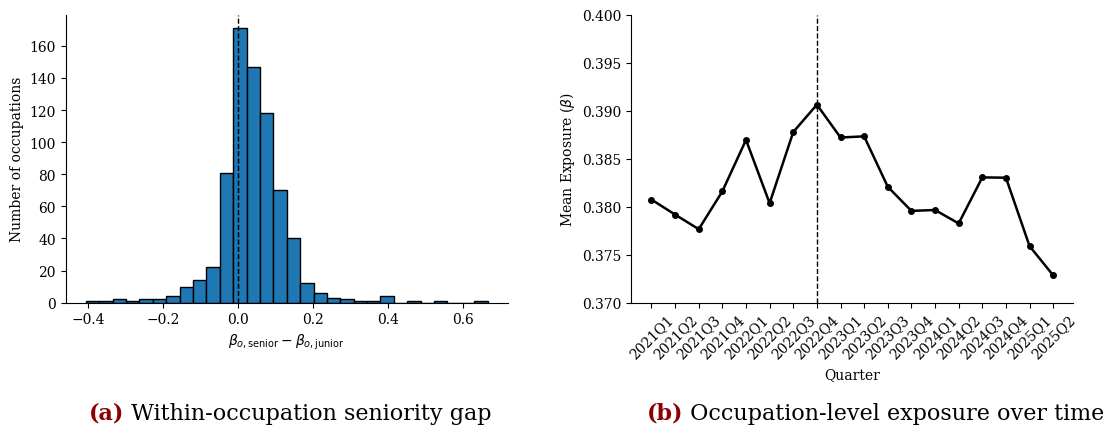}
    \caption{Additional Heterogeneity Captured by the Posting-Based Measure}
    \label{fig:occupation_comparison_appendix}

    \begin{minipage}{1\textwidth}
        \footnotesize
        \textit{Notes}: Panel (a) plots the distribution of $
        \bar{\beta}_{o,\text{senior}} - \bar{\beta}_{o,\text{junior}}$,
        where \(\bar{\beta}_{o,\text{senior}}\) and \(\bar{\beta}_{o,\text{junior}}\) denote the average exposure of senior and junior postings within occupation \(o\), respectively. Positive values indicate that senior postings are more exposed than junior postings within the same occupation. Panel (b) plots occupation-level exposure over time. For each quarter \(t\), we first compute the mean exposure within each occupation \(o\), and then average these occupation-level means across occupations: $
        \bar{\beta}_t = \frac{1}{|\mathcal{O}_t|}\sum_{o \in \mathcal{O}_t} \bar{\beta}_{o,t}$.
        This panel shows that occupation-level exposure evolves over time rather than remaining fixed at a single occupation-specific value.
    \end{minipage}
\end{figure}

Figure~\ref{fig:occupation_comparison_appendix} summarizes these two additional dimensions directly. Panel (a) plots, for each occupation, the difference between the average exposure of senior postings and that of junior postings. The distribution is centered above zero, indicating that senior postings tend to be more exposed than junior postings even within the same occupation. Panel (b) plots the evolution of occupation-level exposure over time, where exposure is first aggregated to the occupation-by-quarter level and then averaged across occupations within each quarter. This time series shows that occupation-level exposure is not fixed, but instead changes gradually over time. Taken together, these results show that our measure is consistent with the benchmark in \citet{eloundou2024gpts} at the aggregate occupation level, while additionally capturing within-occupation heterogeneity by seniority and within-occupation changes over time.

\newpage
\section{Sector-Level Average Generative AI Exposure ($\beta$)}
\label{app:sector_avg_exposure}

\begin{table}[htbp]
\centering
\caption{Sector-Level Average Generative AI Exposure ($\beta$) by Two-Digit NAICS Code}
\vspace{-9pt}
\label{tab:sector_avg_exposure}

\begin{threeparttable}
\setlength{\tabcolsep}{12pt}

\begin{tabular}{lc}
\toprule
Sector & Average Exposure ($\bar{\beta}$) \\
\midrule
Finance and Insurance (52)                                              & 0.584 \\
Professional, Scientific, and Technical Services (54)                   & 0.548 \\
Information (51)                                                        & 0.545 \\
Educational Services (61)                                               & 0.495 \\
Utilities (22)                                                          & 0.483 \\
Public Administration (92)                                              & 0.465 \\
Manufacturing (31--33)                                                  & 0.454 \\
Management of Companies and Enterprises (55)                            & 0.441 \\
Real Estate and Rental and Leasing (53)                                 & 0.435 \\
Mining, Quarrying, and Oil and Gas Extraction (21)                      & 0.398 \\
Construction (23)                                                       & 0.397 \\
Wholesale Trade (42)                                                    & 0.386 \\
Other Services (except Public Administration) (81)                      & 0.378 \\
Administrative and Support and Waste Management Services (56)           & 0.377 \\
Unclassified / Unknown (99)                                             & 0.372 \\
Agriculture, Forestry, Fishing and Hunting (11)                         & 0.350 \\
Health Care and Social Assistance (62)                                  & 0.350 \\
Arts, Entertainment, and Recreation (71)                                & 0.338 \\
Transportation and Warehousing (48--49)                                 & 0.292 \\
Retail Trade (44--45)                                                   & 0.289 \\
Accommodation and Food Services (72)                                    & 0.239 \\
\bottomrule
\end{tabular}

\begin{tablenotes}
\scriptsize
\item \textit{Notes:} Average exposure is computed as the mean of posting-level $\beta$ within each two-digit NAICS sector, pooled across all quarters in the sample. Sectors are sorted in descending order of average exposure.
\end{tablenotes}

\end{threeparttable}
\end{table}

\newpage
\section{Additional Descriptive Figures}
\label{app:add_des_figs}

\begin{figure}[htbp]
    \centering
    \includegraphics[width=\textwidth]{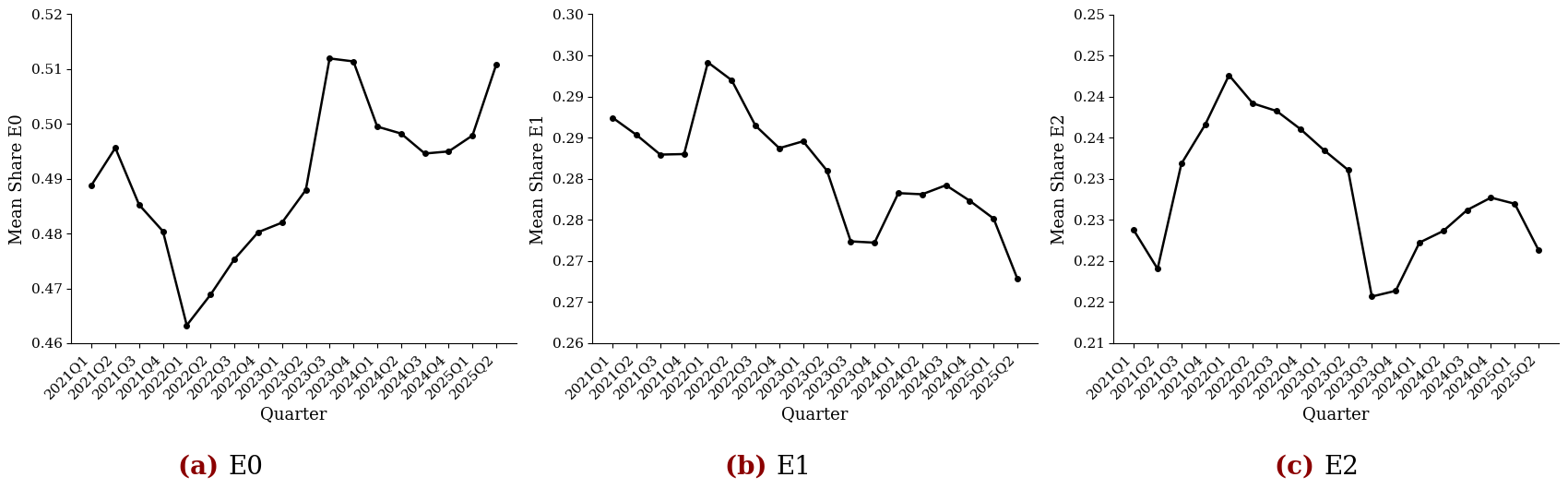}
    \caption{Quarterly Trends in E0, E1, and E2 Shares}
    \label{fig:e0e1e2_appendix}
    \vspace{0.2cm}

    \begin{minipage}{\textwidth}
        \footnotesize
        \textit{Notes}: This figure plots the quarterly mean posting-level weighted shares of task content in E0, E1, and E2 from 2021Q1 to 2025Q2. E0 denotes no meaningful exposure to generative AI, E1 denotes direct exposure, and E2 denotes indirect exposure.
    \end{minipage}
\end{figure}

\begin{figure}[htbp]
    \centering
    \includegraphics[width=\textwidth]{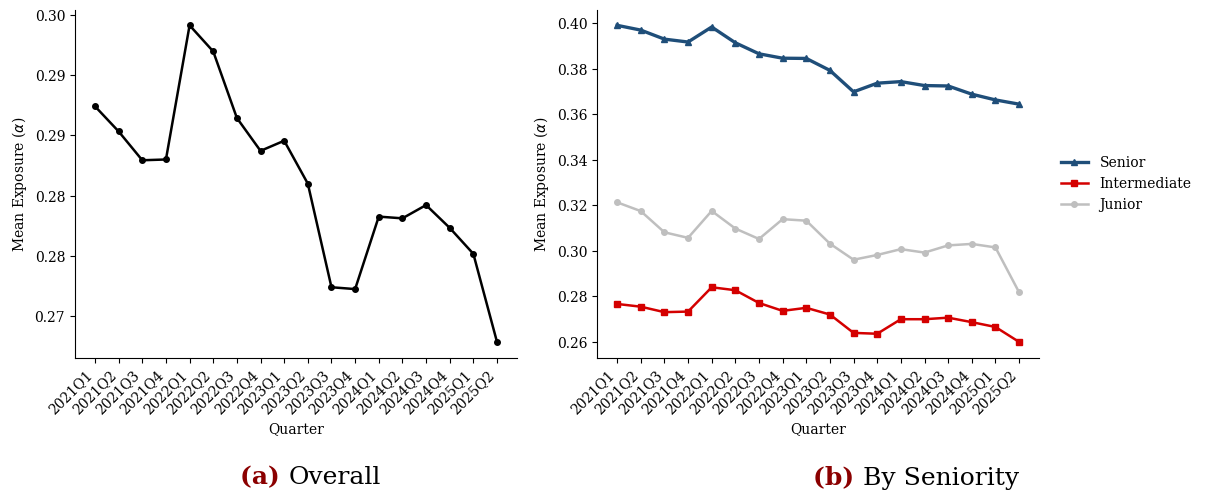}
    \caption{Quarterly Trend in Exposure Measure $\alpha$}
    \label{fig:alpha_appendix}
    \vspace{0.2cm}

    \begin{minipage}{\textwidth}
        \footnotesize
        \textit{Notes}: This figure plots the quarterly trend in the alternative posting-level exposure measure $\alpha$ from 2021Q1 to 2025Q2. As discussed in the main text, $\alpha$ captures only direct exposure.
    \end{minipage}
\end{figure}

\begin{figure}[htbp]
    \centering
    \includegraphics[width=\textwidth]{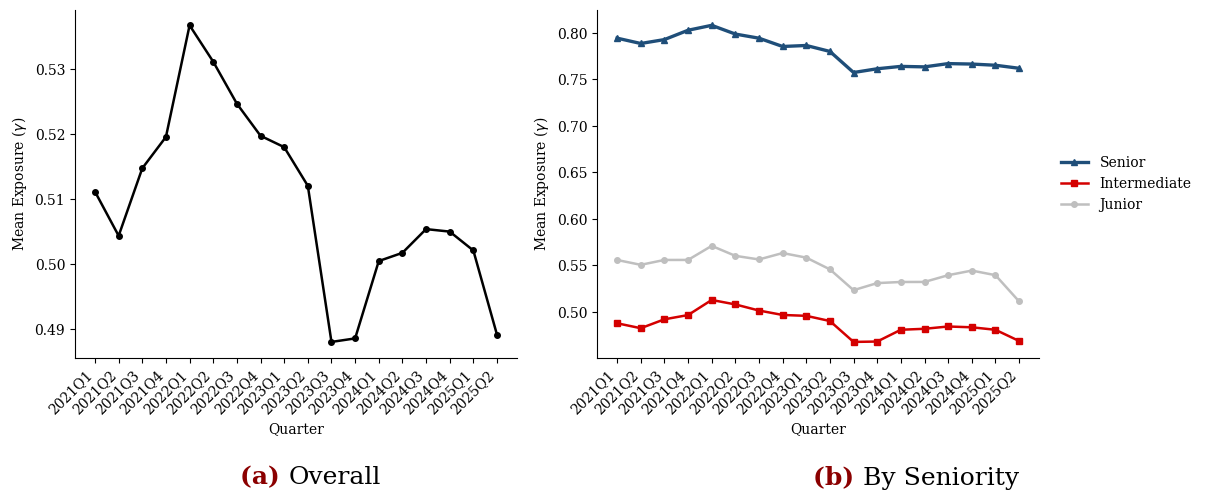}
    \caption{Quarterly Trend in Exposure Measure $\gamma$}
    \label{fig:gamma_appendix}
    \vspace{0.2cm}
    
    \begin{minipage}{1\textwidth}
        \footnotesize
        \textit{Notes}: This figure plots the quarterly trend in the alternative posting-level exposure measure $\gamma$ from 2021Q1 to 2025Q2. As discussed in the main text, $\gamma$ assigns full weight to both direct and indirect exposure.
    \end{minipage}
\end{figure}

\begin{figure}[htbp]
    \centering
    \includegraphics[width=\textwidth]{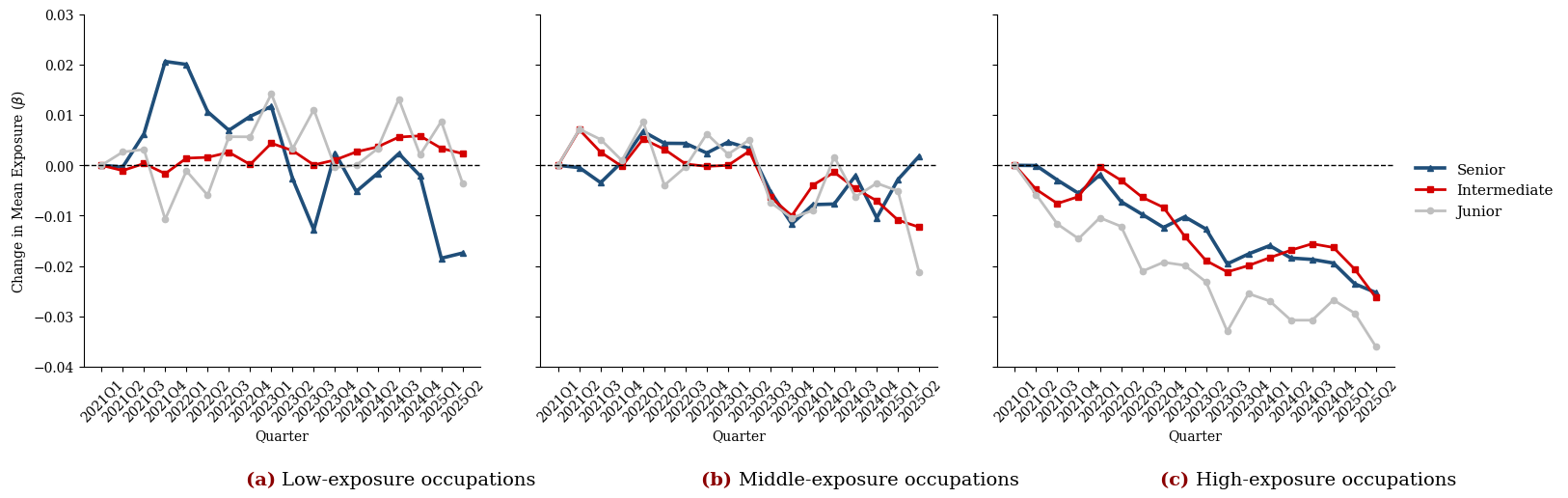}
    \caption{Changes in Generative AI Exposure by Occupation Group and Job Seniority}
    \label{fig:occ_terciles_seniority_appendix}
    \vspace{0.2cm}

    \begin{minipage}{1\textwidth}
        \footnotesize
        \textit{Notes}: This figure groups occupations into three categories based on their average posting-level exposure, $\beta$, over the full sample period: low-, middle-, and high-exposure occupations. Within each occupation group, it plots the quarterly change in mean exposure separately for junior, intermediate, and senior postings. The vertical dashed line marks 2022Q4, and the horizontal dashed line indicates zero change.
    \end{minipage}
\end{figure}

Figure~\ref{fig:e0e1e2_appendix} plots the quarterly shares of postings classified as E0, E1, and E2. Figure~\ref{fig:alpha_appendix} and Figure~\ref{fig:gamma_appendix} report the quarterly trends in the alternative exposure indices $\alpha$ and $\gamma$. Figure~\ref{fig:occ_terciles_seniority_appendix} further shows the quarterly change in $\beta$ by occupation group and job seniority.

\newpage
\section{Occupations with the Highest and Lowest Exposure to Generative AI}
\label{app:top_bottom_occ}

Table~\ref{tab:top_bottom_occ_exposure} presents the ten occupations with the highest and lowest average exposure across the full sample. The results illustrate substantial heterogeneity across occupations. Occupations involving content creation, writing, and digital tasks tend to exhibit high exposure, while occupations involving manual, physical, or routine operational tasks exhibit low exposure.

\begin{table}[htbp]\centering
\caption{Top and Bottom Occupations by Generative AI Exposure}
\vspace{-9pt}
\label{tab:top_bottom_occ_exposure}
\begin{threeparttable}
\begin{tabular}{lllp{7cm}r}
\toprule
Group & Rank & O*NET Code & Occupation & Mean Exposure ($\beta$) \\
\midrule
\multicolumn{5}{l}{\textit{Bottom 10 Occupations}} \\
\midrule
 & 1 & 51-6021.00 & Pressers, Textile, Garment, and Related Materials & 0.057 \\
 & 2 & 47-3012.00 & Helpers--Carpenters & 0.052 \\
 & 3 & 35-2015.00 & Cooks, Short Order & 0.043 \\
 & 4 & 19-4051.00 & Nuclear Technicians & 0.042 \\
 & 5 & 35-9021.00 & Dishwashers & 0.041 \\
 & 6 & 53-7081.00 & Refuse and Recyclable Material Collectors & 0.035 \\
 & 7 & 49-9097.00 & Signal and Track Switch Repairers & 0.023 \\
 & 8 & 51-9195.03 & Molders, Shapers, and Casters, Except Metal and Plastic & 0.018 \\
 & 9 & 53-6032.00 & Aircraft Cargo Handling Supervisors & 0.010 \\
 & 10 & 47-2072.00 & Pile-Driver Operators & 0.000 \\
\midrule
\multicolumn{5}{l}{\textit{Top 10 Occupations}} \\
\midrule
 & 1 & 27-3043.05 & Poets, Lyricists and Creative Writers & 0.839 \\
 & 2 & 27-3043.00 & Writers and Authors & 0.782 \\
 & 3 & 27-3042.00 & Technical Writers & 0.777 \\
 & 4 & 15-1254.00 & Web Developers & 0.749 \\
 & 5 & 13-1199.04 & Online Merchants & 0.741 \\
 & 6 & 19-3099.00 & Social Scientists and Related Workers, All Other & 0.739 \\
 & 7 & 15-1299.06 & Search Marketing Strategists & 0.727 \\
 & 8 & 19-3094.00 & Political Scientists & 0.727 \\
 & 9 & 27-3041.00 & Editors & 0.724 \\
 & 10 & 25-1082.00 & Library Science Teachers, Postsecondary & 0.718 \\
\bottomrule
\end{tabular}
\begin{tablenotes}
\small
\item Notes: This table reports the occupations with the highest and lowest average exposure to generative AI, measured by $\beta$, across the full sample of job postings. Exposure is computed at the posting level based on task-level classifications and then aggregated to the occupation level. Occupations are defined using O*NET codes.
\end{tablenotes}
\end{threeparttable}
\end{table}

\newpage
\section{Common Support and Renormalization}
\label{app:common_support}

A practical issue in the decomposition is that the set of observed occupation $\times$ seniority $\times$ industry cells changes over time. Some cells are observed in both the 2021 baseline and period \(t\), while others appear in only one of the two periods. This support non-overlap can mechanically affect the decomposition, especially in the three-fold specification, where part of the change may be absorbed by the interaction term.

To address this issue, we define the period-specific common support between 2021 and period \(t\) as
\begin{equation*}
S_t = \{ c : w_{c,2021} > 0 \text{ and } w_{ct} > 0 \}.
\end{equation*}
Thus, \(S_t\) contains only cells that are observed in both the baseline and the current period.

After restricting attention to \(S_t\), we renormalize posting shares within the common-support sample. Specifically, for each \(c \in S_t\), define
\begin{equation*}
\tilde{w}^{(t)}_{ct}
=
\frac{w_{ct}}{\sum_{j \in S_t} w_{jt}},
\qquad
\tilde{w}^{(t)}_{c,2021}
=
\frac{w_{c,2021}}{\sum_{j \in S_t} w_{j,2021}}.
\end{equation*}
These renormalized weights sum to one within the common support in each period:
\begin{equation*}
\sum_{c \in S_t} \tilde{w}^{(t)}_{ct} = 1,
\qquad
\sum_{c \in S_t} \tilde{w}^{(t)}_{c,2021} = 1.
\end{equation*}

Using these weights, the common-support aggregate exposure in period \(t\) is
\begin{equation*}
\bar{E}^{CS,\;renorm}_t
=
\sum_{c \in S_t} \tilde{w}^{(t)}_{ct} E_{ct},
\end{equation*}
and the corresponding baseline object is
\begin{equation*}
\bar{E}^{CS,\;renorm}_{2021}(t)
=
\sum_{c \in S_t} \tilde{w}^{(t)}_{c,2021} E_{c,2021}.
\end{equation*}

Renormalization is useful because, without it, a decomposition on the common support would still reflect not only changes among persistent cells, but also changes in how much total posting mass lies on the common support. By renormalizing, we isolate the change among cells that are observed in both periods.

The renormalized common-support object is not, in general, identical to the raw aggregate exposure. We therefore report the difference between the raw aggregate and the renormalized common-support aggregate as a summary measure of support non-overlap:
\begin{equation*}
\text{Gap}_t = \bar{E}_t - \bar{E}^{CS,\;renorm}_t.
\end{equation*}
A similar comparison can be made for the baseline period using \(\bar{E}^{CS,\;renorm}_{2021}(t)\).

We also report two simple overlap diagnostics:
\begin{equation*}
m^{cur}_t = \sum_{c \in S_t} w_{ct},
\qquad
m^{base}_t = \sum_{c \in S_t} w_{c,2021}.
\end{equation*}
Here, \(m^{cur}_t\) is the share of current-period posting mass that lies on the common support, and \(m^{base}_t\) is the corresponding share for the 2021 baseline. These measures provide a transparent summary of how much of the posting distribution is comparable across the two periods.

\subsection{Diagnostic results for the renormalized common-support decomposition}
\label{app:diag_main}

Table~\ref{tab:diag_main} reports diagnostic results for the renormalized common-support decomposition used in the main text. For each quarter, the table reports the raw total change in aggregate exposure, the corresponding renormalized common-support total, and the residual difference between the two, together with the composition, within-cell, and interaction components of the renormalized decomposition. The residual remains small in magnitude throughout, though it becomes somewhat more visible around the mid-2023 turning point. The numerical reconstruction gap is negligible in all quarters, confirming that the renormalized total is exactly accounted for by the three-fold decomposition.

\newpage
\begin{table}[htbp]
\centering
\caption{Diagnostic Decomposition Results under Renormalized Common Support}
\vspace{-9pt}
\label{tab:diag_main}
\begin{threeparttable}
\setlength{\tabcolsep}{34pt}
\begin{tabular}{lrrrrrr}
\toprule
Quarter & Raw total & Renorm.\ total & Residual \\
\midrule
2022Q1 &  0.0169 &  0.0170 & -0.0001  \\
2022Q2 &  0.0130 &  0.0130 &  0.0000 \\
2022Q3 &  0.0071 &  0.0071 & -0.0001  \\
2022Q4 &  0.0032 &  0.0033 & -0.0000 \\
2023Q1 &  0.0028 &  0.0030 & -0.0002 \\
2023Q2 & -0.0020 & -0.0017 & -0.0003  \\
2023Q3 & -0.0183 & -0.0179 & -0.0005  \\
2023Q4 & -0.0181 & -0.0177 & -0.0004 \\
2024Q1 & -0.0091 & -0.0089 & -0.0002  \\
2024Q2 & -0.0086 & -0.0083 & -0.0003  \\
2024Q3 & -0.0062 & -0.0060 & -0.0002  \\
2024Q4 & -0.0074 & -0.0071 & -0.0002  \\
2025Q1 & -0.0099 & -0.0097 & -0.0002 \\
2025Q2 & -0.0200 & -0.0196 & -0.0004 \\
\bottomrule
\end{tabular}
\begin{tablenotes}
\scriptsize
\item \textit{Notes:} Raw total is the change in aggregate exposure relative to the 2021 baseline in the original data. Renorm.\ total is the corresponding change computed on the period-specific common-support sample after renormalizing posting shares. Residual is the difference between the raw total and the renormalized total. By construction, the renormalized total is exactly reconstructed by the sum of the composition, within-cell, and interaction effects; the numerical reconstruction gap is negligible throughout.
\end{tablenotes}
\end{threeparttable}
\end{table}

\newpage
\section{Additional Results: Symmetric Kitagawa Decomposition and the balanced-cell Sample }
\label{app:additional_decomp}

This appendix reports two additional sets of decomposition results. First, we present results based on the symmetric two-fold Kitagawa decomposition \citep{kitagawa1955components}. Second, we restrict the sample to balanced job cells that are observed in every quarter and re-estimate the decomposition on this balanced-cell sample. The results from both alternative decompositions are consistent with our main findings.

\subsection{Symmetric two-fold Kitagawa decomposition}
\label{app:symmetric_decomp}

Our main specification uses the three-fold counterfactual Kitagawa decomposition in Equation~\ref{eq:threefold_main}, which separates the overall change in aggregate exposure into a composition effect, a within-cell exposure effect, and an interaction effect. As an alternative, we also consider the symmetric two-fold Kitagawa decomposition \citep{kitagawa1955components}, which absorbs the interaction term equally into the composition and within-cell components.

Let aggregate exposure in quarter \(t\) be
\[
\bar{E}_t = \sum_c w_{ct} E_{ct},
\]
where \(w_{ct}\) denotes the posting share of cell \(c\) in quarter \(t\), and \(E_{ct}\) denotes average exposure within that cell. Using 2021 as the fixed baseline, the symmetric two-fold decomposition can be written as

\begin{equation*}
\bar{E}_t - \bar{E}_{0}
=
\underbrace{\sum_c (w_{ct} - w_{c,0}) \frac{E_{ct} + E_{c,0}}{2}}_{\text{Composition effect}}
+
\underbrace{\sum_c (E_{ct} - E_{c,0}) \frac{w_{ct} + w_{c,0}}{2}}_{\text{Within-cell exposure effect}}.
\label{eq:twofold_symmetric}
\end{equation*}

Relative to the three-fold decomposition, this formulation provides a more parsimonious two-part breakdown of the change in aggregate exposure by allocating the interaction term equally to the composition and within-cell exposure effects. However, unlike the three-fold specification, it cannot separately capture the interaction effect, namely the component arising when reallocation across job cells and within-cell exposure changes occur simultaneously.

\begin{figure}[htbp]
    \centering
    \includegraphics[width=\textwidth]{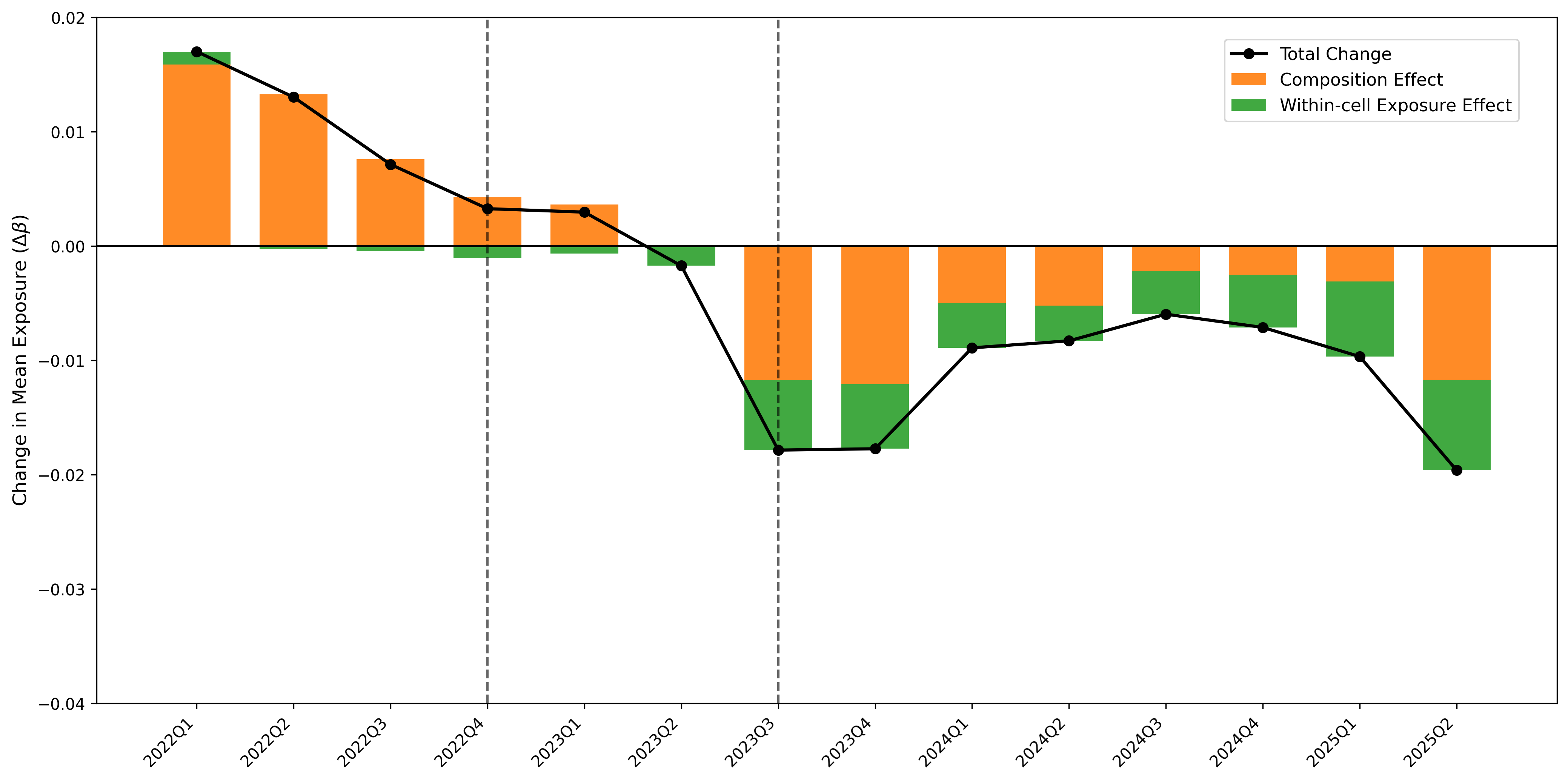}
    \caption{Symmetric Two-Fold Kitagawa Decomposition of Changes in Aggregate Exposure}
    \label{fig:decomp_symmetric_appendix}
\end{figure}

Figure~\ref{fig:decomp_symmetric_appendix} reports the results from the symmetric two-fold decomposition. The pattern is consistent with our main findings: labor-demand adjustment to generative AI operates through both composition shifts and within-cell exposure changes, with both margins becoming more salient after mid-2023.

\subsection{Balanced-cell sample}
\label{app:balanced_cells}

A potential concern in decomposition exercises is that changes in aggregate exposure may partly reflect entry and exit of job cells over time. To address this concern, we re-estimate the decomposition using a balanced-cell sample that includes only cells observed in every quarter of the sample period. We then apply the same three-fold decomposition as in Equation~\ref{eq:threefold_main} to this restricted sample.

\begin{figure}[htbp]
    \centering
    \includegraphics[width=\textwidth]{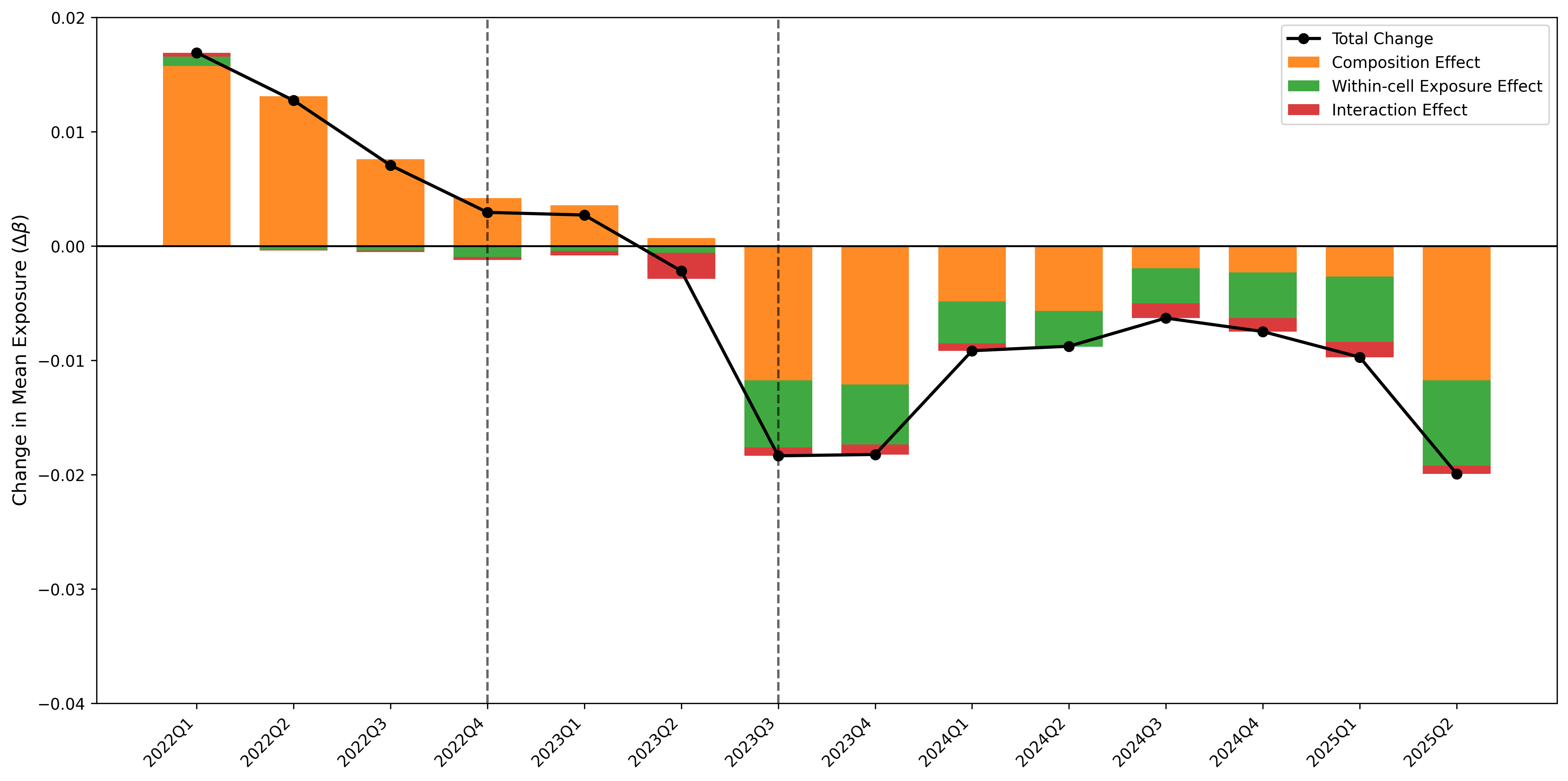}
    \caption{Three-Fold Kitagawa Decomposition Using the Balanced-cell Sample}
    \label{fig:decomp_balanced_appendix}
\end{figure}

Figure~\ref{fig:decomp_balanced_appendix} presents the results based on the balanced-cell sample. The results remain consistent with our main findings. Even when restricting the analysis to cells that are present in every quarter, labor-demand adjustment to generative AI is reflected in both composition shifts and within-cell exposure changes, especially after mid-2023.

\newpage

\newpage
\section{Additional decomposition evidence}

\subsection{Exposure levels and counterfactual paths}
\label{app:overall_level}

\begin{figure}[htbp]
    \centering
    \includegraphics[width=\textwidth]{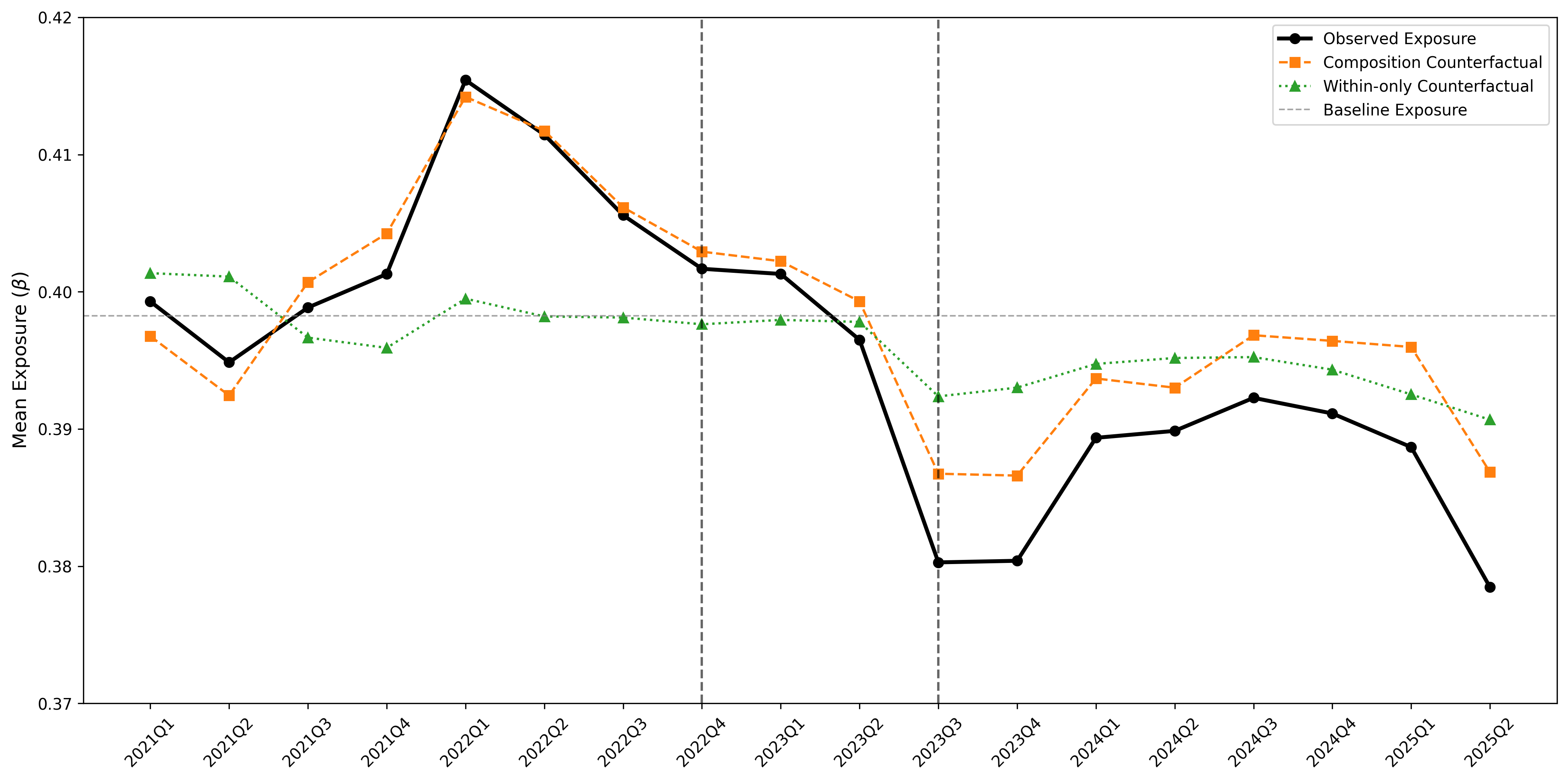}
    \caption{Aggregate Generative AI Exposure over Time and Counterfactual Paths}
    \label{fig:overall_level}
    \vspace{0.2cm}
    
    \begin{minipage}{\textwidth}
        \footnotesize
        \textit{Notes:} The figure plots observed aggregate exposure together with two counterfactual paths. The composition counterfactual allows posting shares to vary over time while holding within-cell exposure fixed at its 2021 level. The within-only counterfactual allows within-cell exposure to vary over time while holding posting shares fixed at their 2021 values. The horizontal dashed line denotes the 2021 baseline exposure.
    \end{minipage}
\end{figure}

Figure~\ref{fig:overall_level} plots observed aggregate generative AI exposure together with two counterfactual paths. The composition counterfactual allows posting shares to vary over time while holding within-cell exposure fixed at its 2021 level, whereas the within-only counterfactual allows within-cell exposure to vary over time while holding posting shares fixed at their 2021 values. This figure provides a level-based benchmark for the decomposition reported in the main text.

The figure confirms the same qualitative pattern as Figure~\ref{fig:overall_change}. In the earlier part of the sample, the composition counterfactual tracks observed exposure closely, while the within-only counterfactual remains much closer to the 2021 baseline. This indicates that the early increase in aggregate generative AI exposure is driven mainly by compositional reallocation across job cells. After 2023Q3, however, both counterfactual paths move downward relative to the baseline, indicating that the later decline reflects both a reversal in hiring composition and a reduction in exposure within cells.

\subsection{Relative contribution of decomposition components}
\label{app:relative_contribution}

\begin{figure}[htbp]
    \centering
    \includegraphics[width=\textwidth]{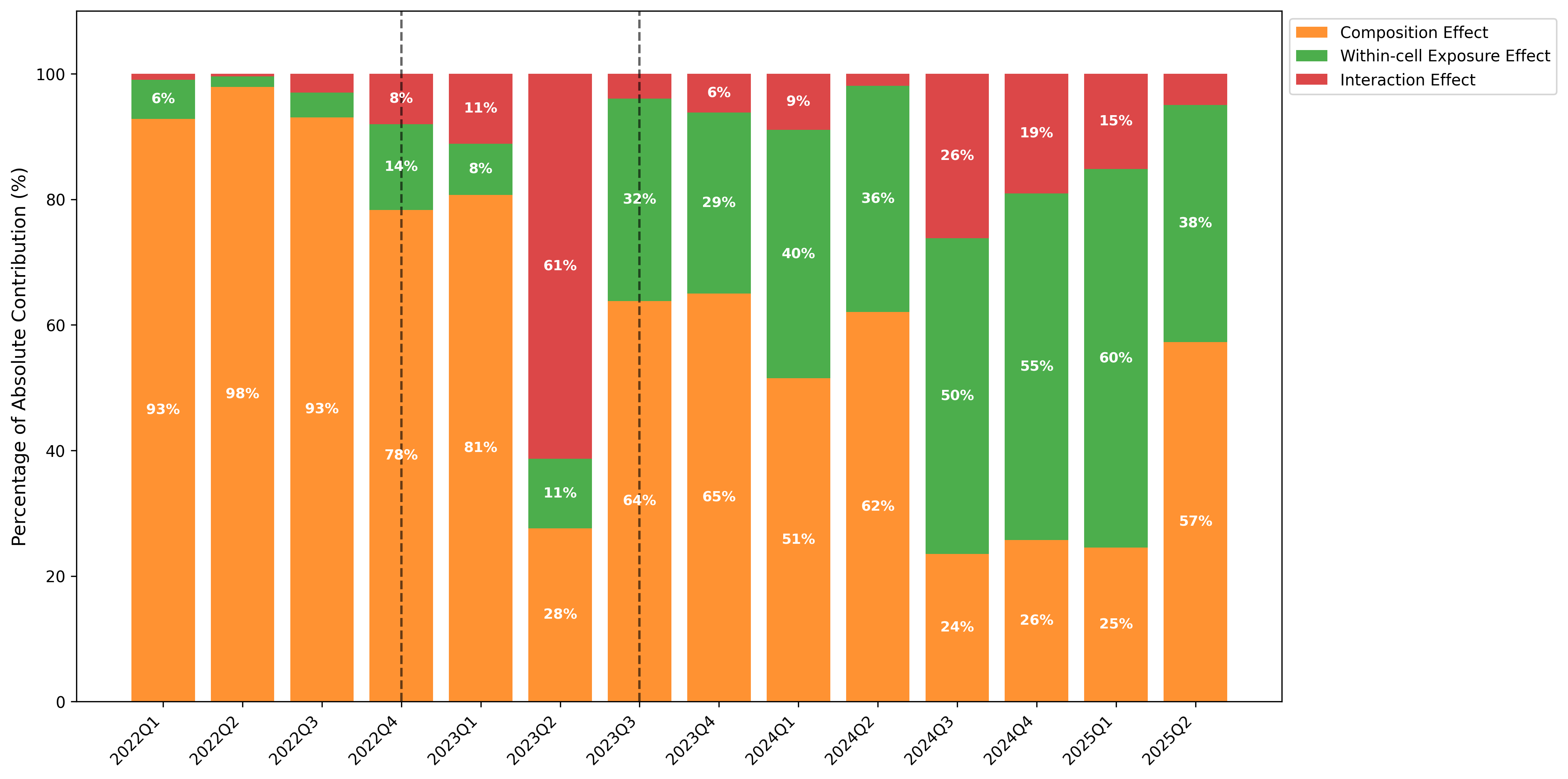}
    \caption{Relative Contribution of Decomposition Components}
    \label{fig:overall_contribution}
    \begin{minipage}{\textwidth}
    \vspace{0.5em}
    \footnotesize
    \textit{Notes:} For each quarter, the figure reports the percentage contribution of each decomposition component in absolute value. The contribution of a component is defined as its absolute value divided by the sum of the absolute values of the composition effect, the within-cell exposure effect, and the interaction effect in that quarter. Quarters begin in 2022Q1 because the decomposition is defined relative to the full-year 2021 baseline.
    \end{minipage}
\end{figure}

Figure~\ref{fig:overall_contribution} reports the relative contribution of each component in absolute terms. For each quarter, the contribution of a component is defined as its absolute value divided by the sum of the absolute values of the composition effect, the within-cell exposure effect, and the interaction effect. This figure complements the main decomposition by showing the relative importance of each margin independent of sign.

The figure reinforces the pattern discussed in the main text. In the earlier part of the sample, the composition effect accounts for the largest share of aggregate generative AI exposure change, consistent with the interpretation that the initial increase in exposure is driven mainly by reallocation in hiring shares across job cells. In the later part of the sample, the relative contribution of the within-cell exposure effect becomes larger, indicating that changes within job cells play an increasingly important role in lowering aggregate exposure. The interaction term remains smaller than the other two components in most quarters, although its contribution rises around some turning points.

\subsection{Unpacking the interaction term}
\label{app:interaction_types_section}

The three-fold decomposition isolates an interaction term that captures the joint movement of hiring shares and within-cell exposure. For cell \(c\), the interaction contribution is given by
\[
\text{Interaction}_{c} = \Delta w_c \times \Delta \beta_c,
\]
where \(\Delta w_c\) denotes the change in the posting share of cell \(c\) relative to the baseline, and \(\Delta \beta_c\) denotes the change in exposure within that cell.

This interaction term can arise from four sign combinations. First, when \(\Delta w_c<0\) and \(\Delta \beta_c>0\), hiring falls in cells whose exposure is rising; this generates a negative interaction contribution. Second, when \(\Delta w_c>0\) and \(\Delta \beta_c<0\), hiring rises in cells whose exposure is falling; this also generates a negative interaction contribution. Third, when \(\Delta w_c>0\) and \(\Delta \beta_c>0\), hiring rises in cells whose exposure is also rising; this generates a positive interaction contribution. Fourth, when \(\Delta w_c<0\) and \(\Delta \beta_c<0\), hiring falls in cells whose exposure is also falling; this likewise generates a positive interaction contribution.

\begin{figure}[htbp]
    \centering
    \includegraphics[width=\textwidth]{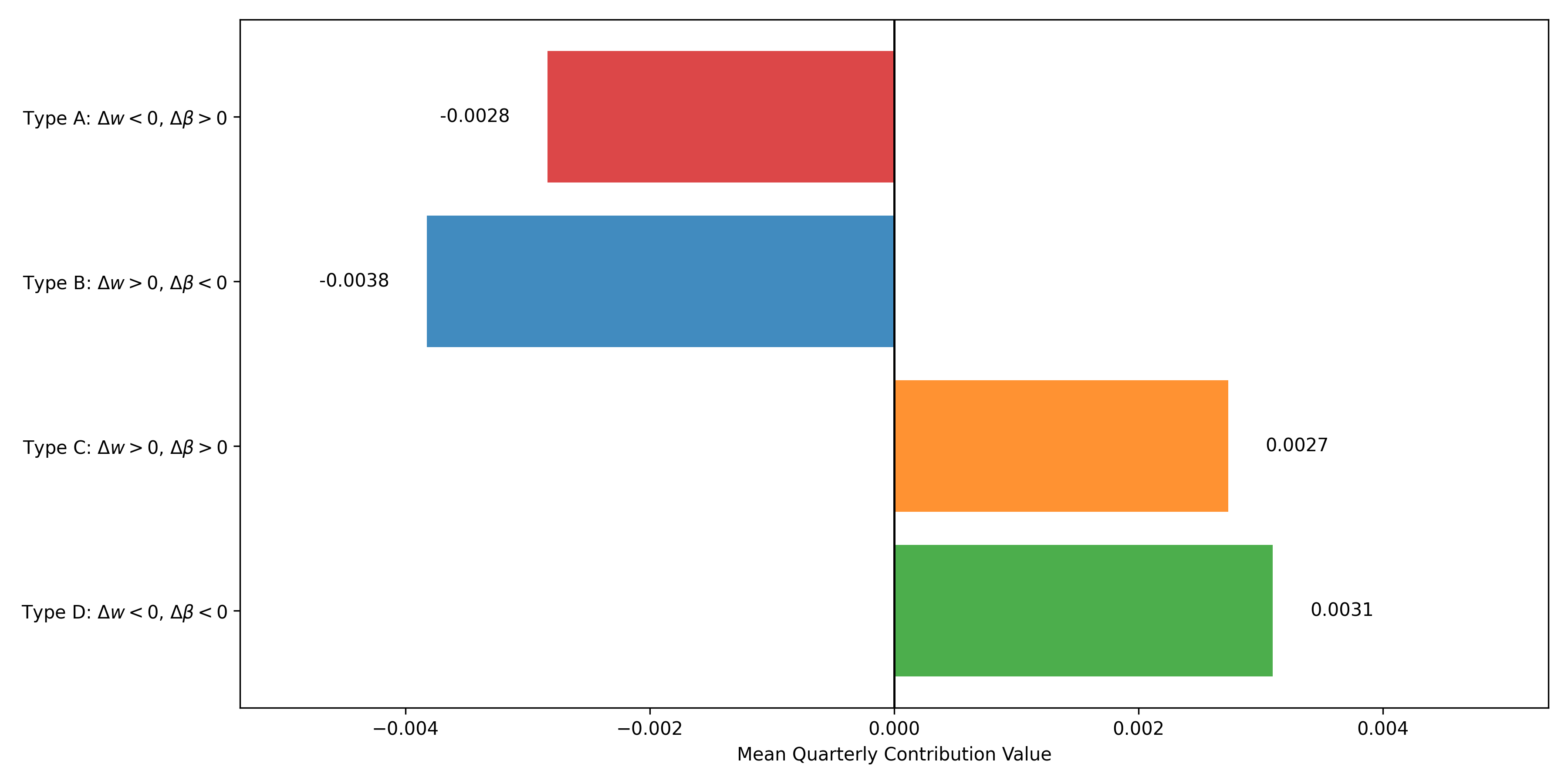}
    \caption{Average Quarterly Contribution to The Interaction Term by Sign Pattern}
    \label{fig:interaction_types}
\end{figure}

Figure~\ref{fig:interaction_types} summarizes the average quarterly contribution of these four cases. The figure shows that the interaction term reflects offsetting patterns across cells rather than the absence of interaction altogether.

\subsection{Sign structure of the interaction term for junior postings}
\label{app:junior_interaction}

\begin{figure}[htbp]
    \centering
    \includegraphics[width=\textwidth]{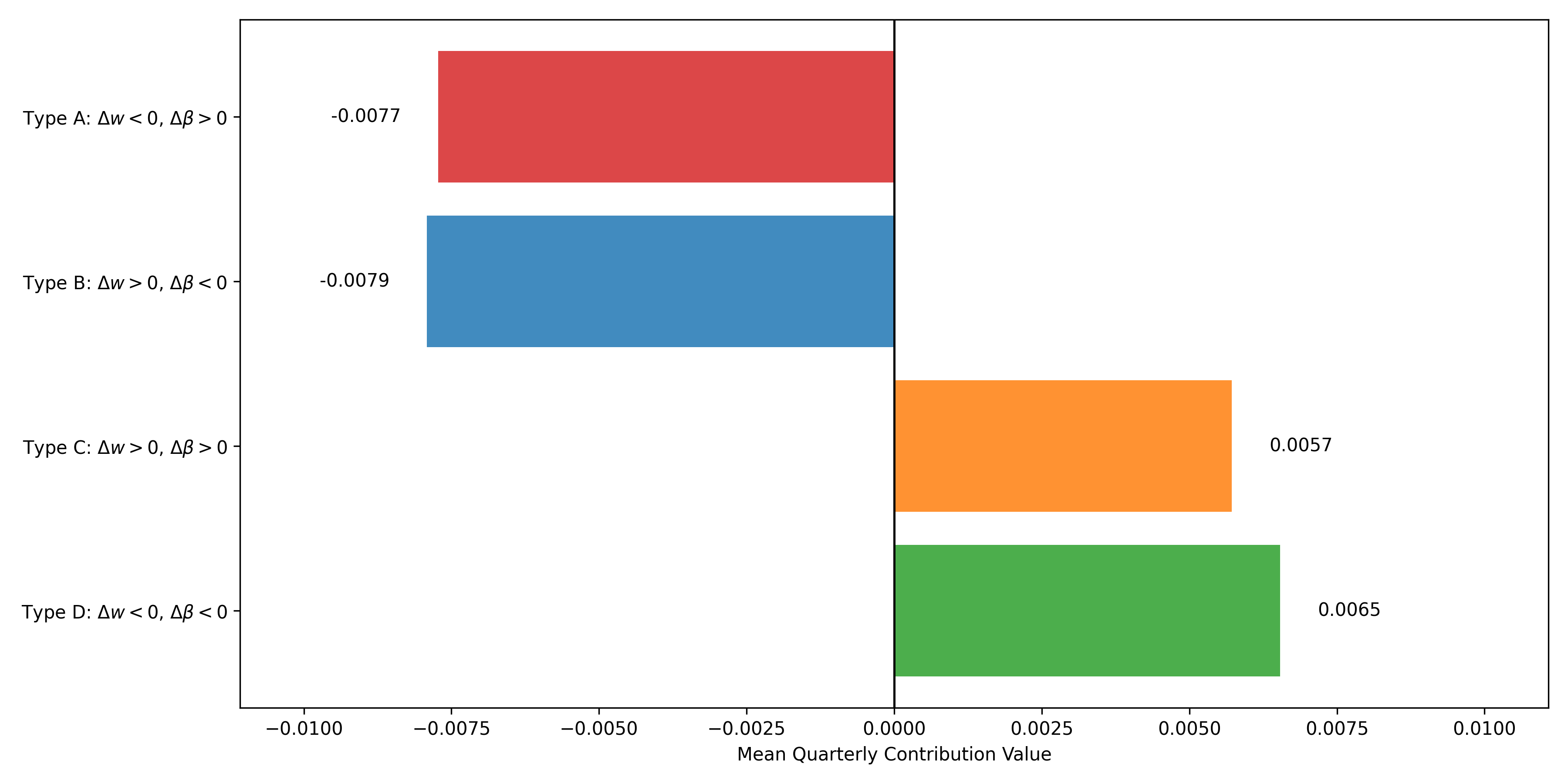}
    \caption{Average Quarterly Contribution to The Interaction Term by Sign Pattern: Junior Postings}
    \label{fig:junior_interaction_types}
\end{figure}

Figure~\ref{fig:junior_interaction_types} decomposes the interaction term for junior postings by the sign of changes in hiring shares and within-cell exposure.

\newpage
\section{Within-Sector Decomposition}
\label{app:within_sector}

One concern is that changes in aggregate exposure may partly reflect sector-level labor demand cycles rather than adjustment to generative AI. This concern is particularly relevant in the post-2022 period, when monetary tightening may have affected hiring across sectors through changes in financing costs, capital expenditures, and demand conditions. Following the logic in \citet{iscenko2026looking}, such macroeconomic shocks are most likely to affect aggregate posting composition through sector-level hiring shifts. For example, if higher interest rates reduce hiring in more interest-sensitive sectors, such as information, finance, or professional services, and these sectors also have higher baseline generative AI exposure, the aggregate composition effect may partly capture cross-sector reallocation rather than AI-related adjustment.

To address this concern, we implement a within-sector version of the decomposition. Specifically, we run the decomposition separately within each two-digit NAICS sector, using variation only across job cells within the same sector. We then aggregate the sector-specific decomposition components using fixed baseline sector weights. This procedure removes cross-sector reallocation as a source of variation: changes in the relative size of sectors over time no longer contribute mechanically to the aggregate decomposition. Instead, the estimates capture whether hiring shifts across occupation-seniority cells within sectors, and whether exposure changes within those cells.

The identifying logic is that, conditional on sector, sector-level macroeconomic shocks are absorbed as a common background shock to postings in that sector. For instance, if higher interest rates reduce total hiring in finance by 20 percent, this sector-level contraction affects the level of finance postings but does not by itself generate a change in the relative composition of finance postings across occupation-seniority cells. The within-sector decomposition therefore asks whether the same adjustment patterns remain after removing the cross-sector channel through which interest-rate shocks are most likely to affect aggregate exposure.

\begin{figure}[htbp]
    \centering
    \includegraphics[width=\textwidth]{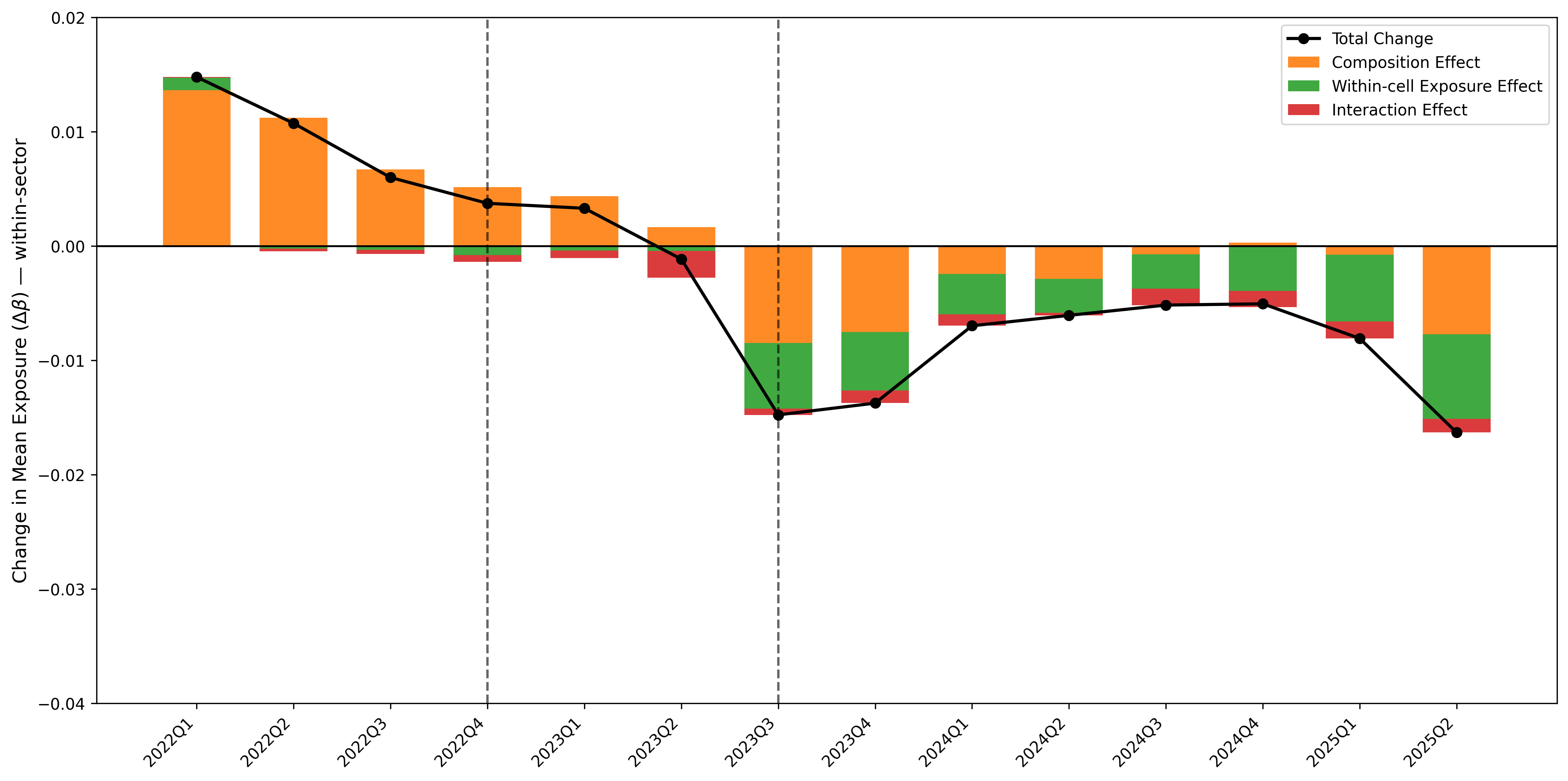}
    \caption{Within-Sector Decomposition of Changes in Generative AI Exposure}
    \label{fig:within_sector}
    \vspace{0.2cm}
    
    \begin{minipage}{\textwidth}
    \footnotesize
    \textit{Notes}: This figure reports a within-sector version of the decomposition. The decomposition is estimated separately within each two-digit NAICS sector and then aggregated using fixed baseline sector weights. 
    \end{minipage}
\end{figure}

Appendix Figure~\ref{fig:within_sector} reports the results. The overall pattern is highly similar to the baseline decomposition. The within-cell exposure effect remains similar in both magnitude and timing, suggesting that the post-2023 decline in within-cell exposure is not driven by cross-sector reallocation. The composition effect is modestly smaller once cross-sector variation is removed, but it remains economically meaningful and follows the same temporal pattern: it is positive before 2023Q3 and turns negative afterward. Thus, sector-level hiring dynamics associated with macroeconomic conditions explain only part of the compositional shift and do not alter the conclusion that labor demand adjusts through both hiring reallocation and within-cell exposure changes.

\end{document}